\documentclass[nofootinbib,amsmath,twocolumn,notitlepage,preprintnumbers,floatfix]{revtex4-1}
\usepackage{multirow}
\usepackage{amscd, amsfonts, amsmath, amssymb, amsthm, eso-pic, esvect,  graphicx, latexsym, physics}
\usepackage{xcolor}
\usepackage{hyperref}
\usepackage{graphicx}
\usepackage{subcaption}
\usepackage{bm}
\usepackage{float}
\usepackage{mathrsfs}
\usepackage{colortbl}
\usepackage{siunitx} 
\usepackage{subfiles}
\usepackage[english]{babel}
\usepackage[justification=raggedright,singlelinecheck=false]{caption}
\usepackage{cleveref}
\usepackage{mathtools}
\usepackage[normalem]{ulem}
\usepackage{lipsum}

\hypersetup{
    colorlinks=true,
    linkcolor=blue,
    citecolor=magenta,      
    urlcolor=cyan}

\newcommand{\eV}{\mathinner{\mathrm{eV}}}
\newcommand{\keV}{\mathinner{\mathrm{keV}}}

\newcounter{num}

\NewDocumentEnvironment{alignlast}{b}{%
  \begin{align*}
  \refstepcounter{equation} #1 \tag{\theequation}
  \end{align*}
}{\ignorespacesafterend}

\def\atag{\refstepcounter{equation}\tag{\arabic{equation}}}
\begin{document}

\title{Probing Heavy Axion-like Particles from Massive Stars with $X$-rays and Gamma Rays}

\author{James H. Buckley$^a$}
\thanks{buckley@wustl.edu}
\author{P.~S.~Bhupal Dev$^a$}
\thanks{bdev@wustl.edu}
\author{Francesc Ferrer$^a$}
\thanks{ferrer@wustl.edu}
\author{Takuya Okawa$^{a,b}$}
\thanks{o.takuya@wustl.edu}

\medskip

\affiliation{$^a$ Department of Physics and McDonnell Center for the Space Sciences, Washington University, Saint Louis, MO 63130, USA}
\affiliation{$^b$ Theoretical Physics Department, Fermi National Accelerator Laboratory, Batavia, IL 60510, USA}

\date{\today}

\begin{abstract}
	The hot interiors of
	massive stars in the later stages of their evolution provide an ideal place for the production of heavy axion-like particles (ALPs) with mass up to $\mathcal{O}(100\ \keV)$ range. We show that a fraction
	of these ALPs could stream out of the stellar photosphere and subsequently decay
	into two photons that can be potentially detected on or near the Earth. 
	In particular, we estimate the photon flux originating from the spontaneous decay of heavy ALPs 
	produced inside Horizontal Branch and Wolf-Rayet
	stars, and assess its detectability by current and future $X$-ray and gamma-ray telescopes.
	Our results indicate that current and future telescopes can probe
	axion-photon couplings down to $g_{a\gamma} \sim 4\times 10^{-11}$ 
	GeV${}^{-1}$ for $m_a\sim 10-100$ keV, which covers new ground in the ALP parameter space. 
\end{abstract}

\maketitle

\section{Introduction}
\label{sec: introduction}

The QCD axion is the pseudo Nambu-Goldstone (pNGB) boson associated with the 
Peccei-Quinn (PQ) symmetry, that provides an elegant solution to the strong CP 
problem~\cite{Peccei:1977hh,Weinberg:1977ma,Wilczek:1977pj} and can also account for the dark matter in the Universe~\cite{Preskill:1982cy,Abbott:1982af,Dine:1982ah}.
More generally, axion-like-particles (ALPs) appear as pNGBs in theories with
a spontaneously broken global $U(1)$ symmetry unrelated to the PQ mechanism,
and are ubiquitous in string-inspired extensions of the Standard 
Model (SM)~\cite{Svrcek:2006yi,Arvanitaki:2009fg}. ALPs are predicted to
have interactions with SM particles analogous to those of QCD axions, but
the latter are restricted to couple in inverse proportion to their mass, whereas the ALP mass and couplings can be treated as independent parameters.
The ALP parameter space is being probed by numerous laboratory and astrophysical
searches~\cite{Choi:2020rgn}.

The original QCD axion had a keV--MeV mass and electroweak scale interactions. 
It was soon experimentally ruled out~\cite{Donnelly:1978ty}, leading to the 
emergence of \textit{invisible} QCD axion 
models~\cite{Kim:1979if,Shifman:1979if,Zhitnitsky:1980tq,Dine:1981rt,Rubakov:1997vp,Berezhiani:2000gh}. 
Although the QCD axion that addresses the strong CP problem is predicted to have 
a mass lighter than $\lesssim \si{eV}$, `heavy' ALPs appear in many 
ultraviolet (UV) completions of the SM~\cite{DiLuzio:2020wdo}, e.g. as decay products of dark matter candidates~\cite{Nomura:2008ru}.

Interestingly, heavy axions with masses 
$m_a \gtrsim \mathcal{O}(10\si{\ keV})$ can have 
lifetimes $\tau_a$ comparable to the age of the Universe 
$t_U \sim 4 \times 10^{17}\si{\ s}$, i.e. 
\begin{align}
    \tau_a = \frac{64 \pi}{g_{a\gamma}^2 m_a^3 } \simeq 10^{17} \si{\ s}\left(\frac{10^{-12}\si{\ GeV^{-1}}}{g_{a\gamma}}\right)^2 \left(\frac{10~\si{keV}}{m_a}  \right)^3,
    \label{eq:lifetime}
\end{align}
where $g_{a\gamma}$ is the axion-photon coupling that induces the 
$a\rightarrow \gamma \gamma$ decay.
Thus, if produced in the late Universe, heavy ALPs typically decay on 
timescales shorter than the age of the Universe and could give rise
to detectable photon signals. For this reason, we will 
focus on ALPs with masses in the range from keV to MeV.

The axion-photon coupling $g _{a\gamma}$ for ALPs in this mass range is constrained by cosmological observations, high energy experiments, as well as astrophysical phenomena. In the context of cosmology, ALPs decaying to photons in the early Universe affect cosmological observables in a model-dependent way, impacting Big-Bang Nucleosynthesis (BBN) and distortions of the Cosmic Microwave Background (CMB) radiation spectrum. 
ALPs can be produced thermally, by the decay of heavier particles or 
from topological defects such as cosmic strings and domain walls, or 
from primordial black holes via the Hawking evaporation process. 
The injection of their decay products into the thermal bath of the early
Universe modifies the abundances of nuclei such as Deuterium and Helium-4, and it can also
induce distortions of the CMB spectrum~\cite{Cadamuro:2011fd,Millea:2015qra,Depta:2020wmr}. 
Moreover, heavy ALPs could be 
produced at colliders: they give rise to the process $e ^{+} e ^{-} \to \gamma ^{*} \to \gamma + a$ at LEP~\cite{Jaeckel:2015jla} and the radiative upsilon decay $\Upsilon \to \gamma ^{*} \to \gamma + a$ at the Crystal Ball~\cite{CrystalBall:1990xec}, BaBar~\cite{BaBar:2008aby} and Belle~\cite{Belle-II:2020jti} experiments. Limits on ALPs from radiative $Z$-boson decays were obtained by the L3 collaboration~\cite{L3:1997exg} and, more recently, by the ATLAS experiment~\cite{ATLAS:2015rsn}. In addition, electron beam dump experiments including E141 and E137~\cite{Bjorken:1988as, Dolan:2017osp}, as well as proton beam dump experiments such as CHARM~\cite{CHARM:1985anb} and NuCal~\cite{Blumlein:1990ay,Blumlein:1991xh} are effective in probing such heavy ALPs.

Heavy ALPs would also be copiously produced in hot and dense astrophysical environments. ALPs produced in supernovae 
could contribute significantly to energy loss
~\cite{Masso:1995tw}, may produce a large flux of photons from decays that could be detected by telescopes~\cite{Jaeckel:2017tud, Ferreira:2022xlw, Hoof:2022xbe,Caputo:2021rux,Muller:2023pip}, or can contribute to the energetics of expanding supernovae~\cite{Caputo:2022mah}. ALPs 
could also be 
produced in main sequence (MS) stars~\cite{Nguyen:2023czp}, horizontal branch (HB) stars~\cite{Ayala:2014pea,Carenza:2020zil,Dolan:2022kul}, Wolf-Rayet (WR) stars~\cite{Dessert:2020lil}, neutron stars~\cite{Berenji:2016jji} and neutron star mergers~\cite{Dev:2023hax,Diamond:2023cto}.

In this paper, we consider the production of heavy ALPs with $m_a \simeq (10-100)$ keV in HB stars and WR stars 
which have high enough core temperatures to produce these particles. 
An HB star consists of a Helium-burning core and a Hydrogen-burning shell. Stars reach the horizontal branch after leaving the MS and going through a red giant phase. On the other hand, WR stars~\cite{abbott1987wolf} are massive stars whose outer envelopes are stripped away because of their high rotational speeds. 
Such stars are also likely to have Helium-burning layers that reach temperatures of $\mathcal{O}(10)$ keV (or $10^8$ K), which is about ten times larger than that of MS stars, and are conducive for the production of heavy ALPs. ALPs produced in the plasma of such stars undergo decays to two photons that might be observable by telescopes on Earth unless the decay occurs inside the photosphere of the star. Previous studies searching for axions produced inside HB stars or WR stars looked for the effect of the star's energy loss~\cite{Ayala:2014pea,Carenza:2020zil,Dolan:2022kul} or $X$-rays converted from light axions~\cite{Dessert:2020lil}. In this paper, we focus on the decaying heavy ALP scenario, and estimate the flux of photons from decays of 
heavy ALPs 
produced in HB stars in two globular clusters, i.e.~NGC 6397 and NGC 2808, and that from WR stars in the Quintuplet Cluster. These particular sources are chosen based on the distance from the Earth (closer is better) and the number of HB stars (more is better). We then discuss their detectability at future $X$-ray/gamma-ray telescopes.

The paper is structured as follows. Sec.~\ref{sec: alp production} briefly 
summarizes the production mechanisms of heavy axions inside stars. The flux of photons at the Earth from their decay is estimated in Sec.~\ref{sec: fluence of photon observed at the Earth}, which also includes a description of the reach of $X$-ray/gamma-ray telescopes and the expected background. Finally, the results and conclusions can be found in Sec.~\ref{sec:results} and Sec.~\ref{sec:conclusion}, respectively. A detailed description of the modeling of WR stars is provided in Appendix~\ref{appendix: WR stars}.

\section{ALP production in plasma}
\label{sec: alp production}

Here we focus on signals that are caused by 
the interaction of ALPs with the electromagnetic field. This is encoded
in the following term in the Lagrangian density:
\begin{align}
    \mathcal{L}_{a\gamma}=-\frac{g_{a \gamma}}{4} F_{\mu \nu} \widetilde{F}^{\mu \nu} a \, ,
\end{align}
where $g _{a\gamma}$ is the ALP-photon coupling, which is taken to be
independent of the mass of the ALP field $a$; $F _{\mu\nu}$ is the 
electromagnetic field-strength tensor, and $\widetilde{F} _{\mu\nu}$ is its 
dual. This interaction term leads to the production of axions in stars via 
two processes: (i) the Primakoff 
process~\cite{Primakoff:1951iae,Cadamuro:2011fd,DiLella:2000dn,Carenza:2020zil}
$\gamma + Ze \to Ze + a$, in which a photon is converted into an axion in the
electrostatic field of charged particles in the plasma, and (ii) photon 
coalescence~\cite{Bastero-Gil:2021oky,Caputo:2022mah,Ferreira:2022xlw}
$\gamma + \gamma \to a$ where two photons in the plasma annihilate into one 
axion. Production rates of axions by these processes have different dependence
on the axion mass $m _{a}$ and the plasma temperature $T$. In the interior
of HB and WR stars, with temperatures reaching $T\sim 10^8$ K, 
axions with mass up to $m _{a} \lesssim 50 \si{\ keV}$ are more efficiently
produced by the Primakoff process. In contrast, those with higher masses,
$m _{a} \gtrsim 50 \si{\ keV}$, are mostly produced by photon 
coalescence~\cite{Carenza:2020zil}. 
We do not consider other  
interaction terms (e.g.~axion-electron or axion-nucleon coupling), and their 
associated production processes, which could in principle enhance the detection prospects discussed here.   

\begin{figure*}[t!]
    \centering
    \begin{subfigure}{.5\textwidth}
      \centering
      \includegraphics[width=.7\linewidth]{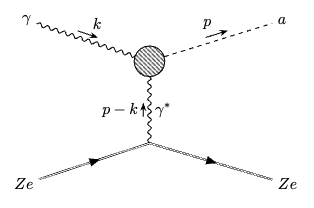}
    \end{subfigure}%
    \begin{subfigure}{.5\textwidth}
      \centering
      \includegraphics[width=.7\linewidth]{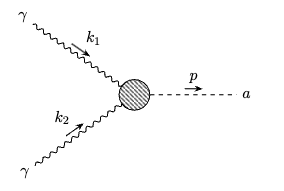}
    \end{subfigure}
    \caption{\label{fig:axion production processes} Two main processes considered here that can efficiently produce axions in the stellar core: the Primakoff process (left) and photon coalescence (right). }
\end{figure*}

In the following, the production spectra (i.e. the number density of axions produced per unit time and frequency) from the Primakoff process and photon coalescence are summarized. The momentum of each particle is labeled as in Fig.~\ref{fig:axion production processes}. In the following, we have only considered the two transverse modes of the photon, which follow the dispersion relation $\omega^2 = k^2 + \omega_{\mathrm{pl}}^2$. Here $k$ is the momentum of a photon, $\omega$ is the frequency of a photon, and $\omega_\mathrm{pl}$ is the plasma frequency. See Sec.~\ref{subsec:thermal} for a discussion of the longitudinal mode contribution.  

\subsection{Primakoff Process}
\label{paragraph: primakoff}
Photons, electrons, and ionized nuclei are in thermal equilibrium in the core of a star. The motion of this plasma induces strong magnetic fields that could convert a photon in the stellar plasma into an axion by the Primakoff process $\gamma(k) + Ze^-(k^\prime) \to a(p) + Ze^-(p^\prime)$. Here $k^\mu = (E_\gamma, \mathbf{k})$ is the four-momentum of the incoming photon, $p = (E_a, \mathbf{p})$ is that of the outgoing axion, $k^\prime = (E_e, \mathbf{k}^\prime)$ is that of the incoming electron/nucleus with atomic number $Z$, $p^\prime = (E_e^\prime, \mathbf{p}^\prime)$ is that of the outgoing nucleus, and $q \equiv p-k = k^\prime - p^\prime$ is the momentum transfer. 

The invariant matrix of this process is given by
\begin{align}
    -i\mathcal{M} = \bar{u}(p^\prime)(iZe\gamma^\alpha)u(k^\prime) \frac{-i\eta_{\alpha\nu}}{q^2} (-ig_{a\gamma}) \epsilon_\mu(k) k_\rho q_\sigma \epsilon^{\mu\nu\rho\sigma},
\end{align}
where $\eta$ is the metric in the Minkowski space and $\epsilon^{\mu\nu\rho\sigma}$ is a totally anti-symmetric tensor. Since the plasma temperature of stars $T \simeq \mathcal{O}(10~\mathrm{keV})$ is typically much lower than nucleus mass $m_N \gtrsim \mathcal{O}(1 \si{\ GeV})$, we could safely approximate that (i) the nuclei are at rest in the rest frame of the star's plasma and (ii) the recoil of the nucleus involved in the Primakoff process is negligible. Correspondingly, we approximate $q^\mu \approx (0, \mathbf{q})$ and $k^{\prime\mu}, p^{\prime\mu} \approx (m_N, \mathbf{0})$. Under these assumptions, the squared invariant amplitude, averaged over the initial two transverse modes of the photon and two spin states of the nucleus, and summed over the final two spin states of the nucleus, is given by
\begin{align}
    \left<|\mathcal{M}|^2\right> &= \frac{8 \pi g_{a\gamma}^2 Z^2 \alpha m_e^2}{q^2}[|\mathbf{k}|^2|\mathbf{p}|^2 - (\mathbf{k}\cdot\mathbf{p})^2],
\end{align}
where $\alpha=e^2/4\pi$ is the fine-structure constant. 
The polarization sum of photons is evaluated as $\sum \epsilon_\alpha^* \epsilon_\beta = -\eta_{\alpha\beta}$. The production spectrum of the Primakoff process is then expressed as:
\begin{alignlast}
    \label{eqn: production rate (Primakoff)}
    \frac{\dd n_a}{\dd t} = & g_\gamma g_N  \int \frac{\dd^3 \mathbf{k}}{(2 \pi)^{3} 2 E_{\gamma}} f_B\left(E_{\gamma}\right)\int \frac{\dd^3 \mathbf{k}^\prime}{(2 \pi)^{3} 2 m_{N}} f_{B/F}\left(E_{k^\prime}\right) \\
    &\times \int \frac{\dd^3 \mathbf{p}}{(2 \pi)^{3} 2 E_{a}} \int \frac{\dd^3 \mathbf{p}^\prime}{(2 \pi)^{3} 2 m_{N}} (1 \pm f_{B/F}\left(E_{p^\prime}\right)) \\
    &\times (2\pi)^4 \delta(E_\gamma - E_a)\delta^{(3)}(\mathbf{k}^\prime - \mathbf{p}^\prime) \frac{1}{S} \left<|\mathcal{M}|^2\right>.
\end{alignlast}
Here, $E_\gamma = \sqrt{k^2 + \omega _{\mathrm{pl}} ^{2}}$ is the photon energy, $\omega _{\mathrm{pl}} \simeq 4\pi\alpha n _{e}/m _{e}$ is the plasma frequency, $g_i$ is the number of degrees of freedom of particle $i$, $f_{B/F} = 1/(e^{(E-\mu)/T} \mp 1)$ denotes the Bose/Fermi distribution, and $S=1$ is the symmetry factor. We neglect the Bose enhancement due to axions as their occupation number is quite low; $f_a \ll 1$. In a plasma, the electric field is effectively shielded because charged particles tend to be surrounded by an oppositely charged cloud. To include this screening effect, we follow the formalism suggested in Ref.~\cite{Raffelt:1985nk} where the factor $1/q^4$ in $\left<|\mathcal{M}|^2\right>$ is replaced by $1/(q^2(q^2+\kappa^2))$. Here $\kappa$ is the screening scale or the inverse of the Debye length $\lambda_D$, which measures the effectiveness of screening of electric fields by charged particles in the plasma. Using the Debye-Huckel approximation, the screening scale $\kappa$ is found as
\begin{align}
    \label{eqn: screening scale}
    \kappa^{2}=\frac{4 \pi \alpha}{T}\left(n_{e}^{\mathrm{eff}}+\sum_{j} Z_{j}^{2} n_{j}^{\mathrm{eff}}\right)
\end{align}
in a nondegenerate medium. 
Here $n_j^{\mathrm{eff}}$ is the effective number density:
\begin{align}
    n_j^{\mathrm{eff}} = g_j \int \frac{\mathrm{d}^{3} \mathbf{p}}{(2 \pi)^{3}} f_{B/F} \left(1\pm f_{B/F}\right),
\end{align}
of a particle with a charge $Z _{j} e$ and degeneracy $g_j$. Note that this replacement allows us to reproduce the form factor of the charge distribution of the plasma that consists of charged particles following the Yukawa potential $V = \frac{Ze}{4 \pi \epsilon_0 r} e^{-\kappa r}$ in the limit of negligible recoil of electrons/nuclei. 

The environment of our interest consists not only of electrons but of other ionized nuclei; the core of HB stars contains ${}^1\mathrm{H}$ and ${}^4\mathrm{He}$ and that of WR stars in the WC phase also have heavier nuclei such as ${}^{12}\mathrm{C}$. In order to take into account contributions from all of these nuclei, we assume that the total Primakoff production rate can be obtained by summing Eq.~\eqref{eqn: production rate (Primakoff)} over all kinds of target nuclei:
\begin{align}
    \frac{\dd n_{a}}{\dd t}\bigg|_\mathrm{total} = \sum_{\mathrm{ions, electrons}} \frac{\dd n_{a}}{\dd t}\bigg|_{\mathrm{ions, electrons}}.
\end{align}
Taking a derivative with respect to $E_a$ and performing integrations, we finally obtain the Primakoff production spectrum:
\begin{align*}
    \label{eqn: the production spectra (Primakoff)}
    &\frac{\dd^2 n_{a}}{\dd E_a \dd t}(T(r), \kappa(r), t, E_a)=\frac{g_{a\gamma}^2 T \kappa^2}{32\pi^3} p k f_B\left(E_{\gamma}\right) \\
    &\times \left\{ \frac{\left[(k+p)^{2}+\kappa^{2}\right]\left[(k-p)^{2}+\kappa^{2}\right]}{4 p k \kappa^{2}} \ln \left[\frac{(k+p)^{2}+\kappa^{2}}{(k-p)^{2}+\kappa^{2}}\right] \right. \\
    &\left.\ \ \ \  -\frac{\left(k^{2}-p^{2}\right)^{2}}{4 k p \kappa^{2}} \ln \left[\frac{(k+p)^{2}}{(k-p)^{2}}\right]-1 \right\}, \atag
\end{align*}
where $T(r)$ is the plasma temperature as a function of the distance $r$ from the star's center. This expression is consistent with the one given in Refs.~\cite{Raffelt:1985nk,Brodsky:1986mi,Raffelt:1987np,Raffelt:1990yz,DiLella:2000dn,Cadamuro:2010cz,Payez:2014xsa,Millar:2021gzs,Ferreira:2022xlw,Muller:2023vjm}. Since nuclei are assumed to be at rest, energy conservation tells that the incoming photon and the outgoing ALP have the same energy,~i.e.~$E_a = \sqrt{p ^{2}+ m _{a} ^{2}} = \sqrt{k ^{2} + \omega _{\mathrm{pl}} ^{2}}$, and thus, momenta $k$ and $p$ can be seen as functions of the axion energy $E_a$.

\begin{figure*}
    \centering   \includegraphics[width=0.8\linewidth]{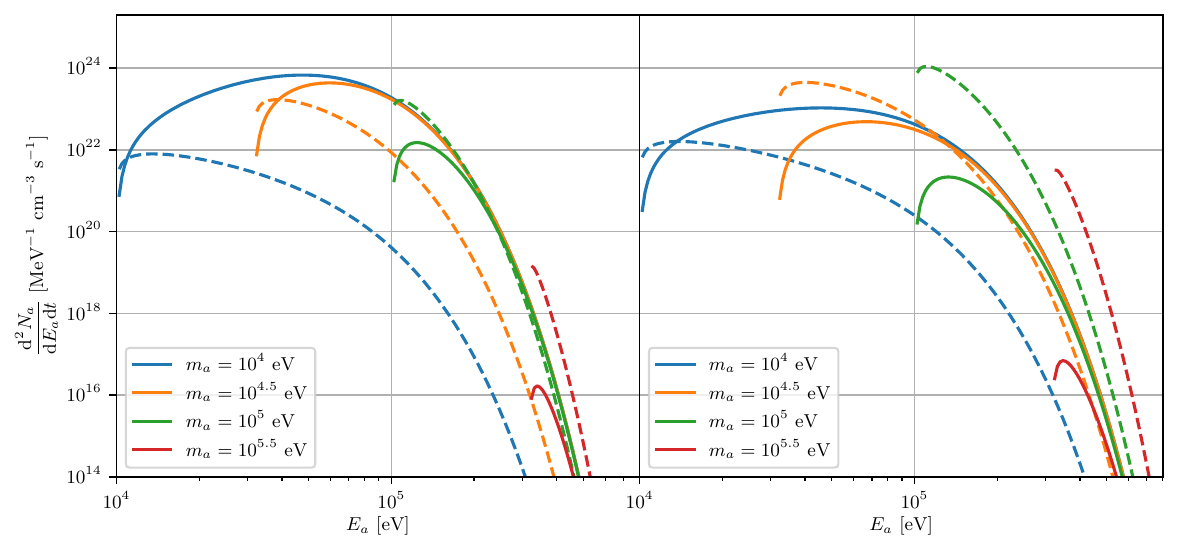}
    \caption{\label{fig: production spectra} Production spectra for the Primakoff process (solid lines) and photon coalescence (dotted lines) for four choices of axion mass: $m_a = 10^4\eV$ (blue), $10^{4.5}\eV$ (orange), $10^5\eV$ (green), and $10^{5.5}\eV$ (orange). In the left panel, we set the values of plasma frequency, temperature, and the inverse Debye length as $\omega_\mathrm{pl} = 3\si{\ keV}, T = 15 \si{\ keV}$, and $\kappa = 50 \si{\ keV}$ which reproduce the environment of the core of a sample HR star (see Fig.~\ref{fig: HB star properties} for details). The right panel assumes WR star-like environment with $\omega_\mathrm{pl} = 3\si{\ keV}, T = 20 \si{\ keV}$, and $\kappa = 5 \si{\ keV}$.}
\end{figure*}
\begin{figure*}
    \centering
    \includegraphics[width=0.8\linewidth]{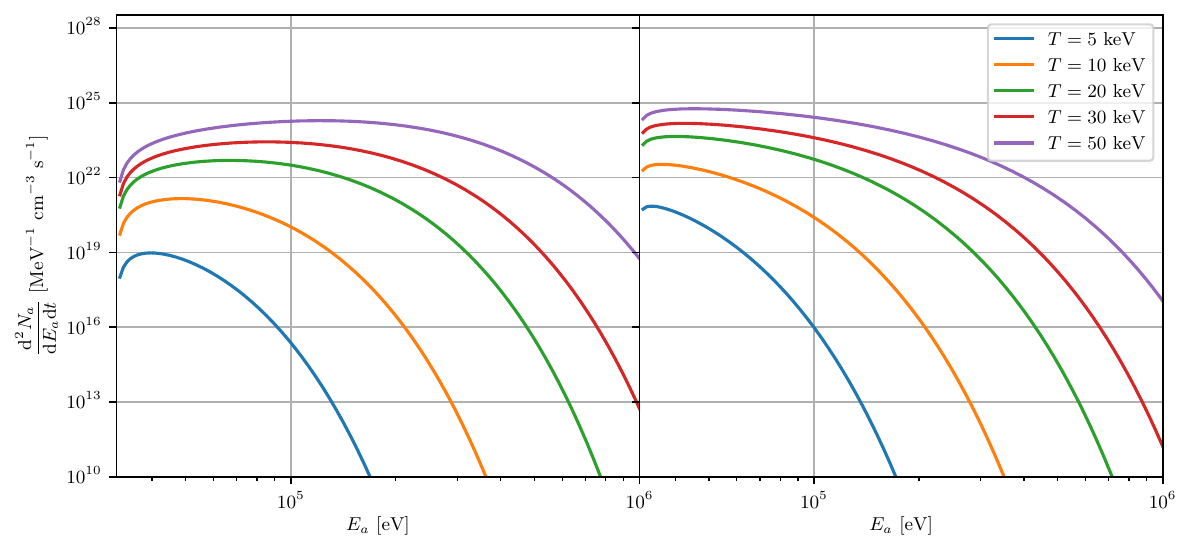}
    \caption{The plot comparing the production spectra of axions in the plasma with different temperatures $T = 5\keV$ (blue), $10\keV$ (orange), $20\keV$ (green), $30\keV$ (red), and $50\keV$ (purple). The axion mass, the plasma frequency, and the axion-photon coupling are fixed to $m_a = 10^{4.5}\si{\ eV}$, $\omega_\mathrm{pl} = 3\si{\ keV}$, and $g_{a\gamma} = 10^{-6}\si{\ GeV^{-1}}$, respectively, i.e. the production rate is calculated at the core of HB star. The left (right) panel shows the production rate of the Primakoff process (photon coalescence).}
    \label{fig:temperature dependence of prodrate}
\end{figure*}

\subsection{Photon Coalescence}
\label{paragraph: photon coalescence}

Axions could also be produced in the stellar plasma via photon coalescence $\gamma(\mathbf{k_1}) + \gamma(\mathbf{k_2}) \to a(\mathbf{p_a})$ in which two photons annihilate into an axion (see Fig.~\ref{fig:axion production processes} right panel). Its production rate is given by 
\begin{align*}
    \label{eqn: production rate (coalescence)}
    \frac{\dd n_a}{\dd t}(r, t)=& \sum \int \frac{\dd^{3} \mathbf{k}_1}{(2 \pi)^{3} 2 E_{1}} f_{1}\left(E_{1}\right) \int \frac{\dd^{3} \mathbf{k}_2}{(2 \pi)^{3} 2 E_{2}} f_{2}\left(E_{2}\right) \\
    &\times \int \frac{\dd^{3} \mathbf{p}_a}{(2 \pi)^{3} 2 E_{a}} (2 \pi)^{4} \delta^{(4)}\left(k_1 + k_2 - p_a\right) \frac{1}{S}  |\mathcal{M}|^{2}, \atag
\end{align*}
where $k_1^\mu = (E_1, \mathbf{k}_1)$, $k_2^\mu = (E_2, \mathbf{k}_2)$, and $p_a^\mu = (E_a, \mathbf{p}_a)$ are four-momenta of the incoming photons (denoted by $1$ and $2$ for convenience) and the outgoing axion. The symmetry factor is $S=2$ for photon coalescence. The summation runs over the spin and polarization states of the photons. The invariant matrix element of this process is given by
\begin{align}
    \mathcal{M} = - i g_{a\gamma} \epsilon^\mu(k_1) \epsilon^\nu(k_2) k_1^\rho k_2^\sigma \epsilon_{\mu\nu\rho\sigma}.
\end{align}
For simplicity, we approximate $s = (k_1 + k_2)^2 \approx m_a^2$ where $s$ is one of the Mandelstam variables. Then, the spin- and polarization-summed squared invariant matrix is reduced to $\left|\mathcal{M}\right|^2=g_{a \gamma}^2 m_a^2\left(m_a^2-4 \omega_\mathrm{pl}^2\right)/2$. Taking a derivative with respect to $E_a$ and performing integrations again gives the production spectrum:
\begin{align*}
    \label{eqn: the production spectra (coalescence)}
    \frac{\dd^2 n_{a}}{\dd E_a \dd t}(r, t, E_a)=& \frac{g_{a\gamma}^2m_a^2}{128\pi^3}(m_a^2 - 4\omega_{\mathrm{pl}}^2) \frac{T}{e^{E_a/T}-1} \\
    &\times\ln\left[ \frac{(1-e^{E_{\max}/T})(e^{E_a/T}-e^{E_{\min}/T})}{(1-e^{E_{\min}/T})(e^{E_a/T}-e^{E_{\max}/T})} \right], \atag
\end{align*}
where $E_{\max}$ ($E_{\min}$) is the maximum (minimum) energy of each photon that can produce an axion with mass $m_a$: 
\begin{align}
E_{\min , \max }=\frac{1}{2}\left(E_a \mp \sqrt{E_a^2 - m_a^2} \sqrt{1-\frac{4 \omega_{\mathrm{pl}}^2}{m_a^2}}\right).
\end{align}
This expression coincides with the result in Ref.~\cite{Carenza:2020zil} in the limit of low temperature compared to the axion mass: $T \ll m_a$.

Fig.~\ref{fig: production spectra} compares the axion production spectra for the Primakoff process (solid lines) and photon coalescence (dashed lines). The left panel assumes the core of HB stars and the right panel assumes that of WR stars. The production spectra are calculated for four benchmark values of axion mass: $m_a = 10^4\eV$ (blue), $10^{4.5}\eV$ (orange), $10^5\eV$ (green), and $10^{5.5}\eV$ (orange). The production spectra are suppressed in the limits of $E_a \to m_a$ due to kinematics and $E_a\gg 3T$ due to Boltzmann suppression. As noted previously, the light axions are produced more efficiently by the Primakoff process while heavier axions are more efficiently produced by photon coalescence. This tendency is more pronounced in a WR star-like environment (right panel). Fig.~\ref{fig:temperature dependence of prodrate} exhibits the axion production spectra for the Primakoff process (left panel) and photon coalescence (right panel) for different plasma temperatures: $T = 5\keV$ (blue), $10\keV$ (orange), $20\keV$ (green), $30\keV$ (red), and $50\keV$ (purple). As expected, more axions are produced in the plasma with higher temperatures. 

\subsection{Thermal Corrections}
\label{subsec:thermal}

In practice, photons acquire a longitudinal mode in plasma~\cite{Weldon:1982aq,Weldon:1983jn,Altherr:1990wi,Altherr:1992mf,Altherr:1993zd,Braaten:1993jw,Kapusta:2006pm}, in addition to the two transverse modes. Let us denote the four-momentum vector of a photon and an electron/positron in the rest frame of the plasma by $K^\mu \equiv (\omega, \mathbf{k})$ and $P \equiv (E, \mathbf{p})$, respectively. The longitudinal mode is described as the mode with the polarization vector whose spatial part is parallel to the wave vector; $\epsilon^\mu_L = (k^2, \omega \mathbf{k})/k\sqrt{K^2}$. One of other three degrees of the photon's polarization vector is taken up by the condition $K^\mu \epsilon_\mu = 0$, which guarantees the polarization tensor of photons $\Pi_{\mu\nu}$ to be transverse; $K^\mu \Pi_{\mu\nu}=K^\nu \Pi_{\mu\nu}=0$. The remaining two degrees of freedom describe transverse modes, i.e. polarization tensors of transverse modes are chosen to be orthogonal to both $K^\mu$ and $\epsilon_L$.

The effect of the finite-temperature plasma is incorporated by calculating the polarization tensor of photons, including the interaction of photons with surrounding electrons and positrons. Summing over all 1-particle irreducible diagrams, one finds the full propagator $\Delta_{\mu\nu}$ where $\Pi_T(\omega,\mathbf{k})$ and $\Pi_L(\omega,\mathbf{k})$ characterize its poles, corresponding to two modes of wave propagation in the plasma. In the Coulomb gauge ($\nabla \cdot \mathbf{A} = 0$), the photon propagator reads as~\cite{An:2013yfc}
\begin{alignlast}
    \Delta_{00} &= \frac{1}{K^2 - \Pi_L}\frac{K^2}{|\mathbf{k}|^2 }, \\
    \Delta_{0i} &= 0, \\
    \Delta_{ij} &= \frac{P_{\mu\nu}}{K^2 - \Pi_T}.
\end{alignlast}
$\Pi_T$ and $\Pi_L$ can be obtained analytically only when $T \gg |\mathbf{k}|, \omega$, and its real part has to do with the dispersion relation while its imaginary part governs the damping behavior of the photon wave. The plasma of stars considered in this paper is nonrelativistic ($T \ll m_e$) and nondegenerate ($m_e -\mu \gg T$). Up to the $\mathcal{O}(T/m_e)$ term, one finds
\begin{align}
    \mathrm{Re}\Pi_T(\omega,\mathbf{k}) &= \omega_p^2\left(1 + \frac{|\mathbf{k}_T|^2}{\omega_T^2}\frac{T}{m_e} \right),\\
    \mathrm{Re}\Pi_L(\omega,\mathbf{k}) &= \omega_p^2\frac{K_L^2}{\omega_L^2}\left( 1 + 3 \frac{|\mathbf{k}_L|^2}{\omega_L^2}\frac{T}{m_e} \right).
\end{align}
At the one-loop level, the contributions from ions to the photon self-energy can be approximately included just by summing over all ions; $\Pi = \sum_{i= e, \mathrm{ions}}\Pi_i$ because photons do not address the internal structure~\cite{Altherr:1993zd}. The real parts remain approximately the same because $m_e \ll m_\mathrm{ions}$. 
The location of the pole of the propagator determines the dispersion relation of each mode:
\begin{align}
    \omega_T^2 &= |\mathbf{k}_T|^2 + \mathrm{Re}\Pi_T(\omega,\mathbf{k}), \\
    \omega_L^2 &= \frac{\omega_L^2}{K_L^2}\mathrm{Re}\Pi_L(\omega,\mathbf{k}).
\end{align}

Insertions of thermal photon diagrams can be interpreted as the modification of the photon self-energy, and thus, the renormalization of the photon field is required. In the finite temperature plasma, the wave function of the photon field associated with the different frequencies is modified differently in a temperature-dependent way. As a function of the frequency and the temperature of the plasma, the wave-function renormalization constant is given by
\begin{align*}
    Z_T & = \left[1 -  \frac{\partial \Pi_T}{\partial \omega_T^2}\right]^{-1} = 1 - \frac{\omega_P^2|\mathbf{k}_T|^2}{\omega_T^4}\frac{T}{m_e}, \atag \\
    Z_L & = \frac{|\mathbf{k}_L|^2}{\omega_L^2}\left[ - \frac{\partial (|\mathbf{k}_L|^2\Pi_L/K_L^2)}{\partial \omega_L^2}\right]^{-1} \nonumber \\
    & = \frac{\omega_L^2}{\omega_P^2}\left(1 - 6 \frac{|\mathbf{k}_L|^2}{\omega_L^2}\frac{T}{m_e}\right),\atag
\end{align*}
again up to the order of $T/m_e$.

We checked that (i) the thermal correction is negligible in our case mainly because of the small plasma temperature compared to the electron mass: $T/m_e \ll 1$, (ii) in photon coalescence, the annihilation of the transverse mode photons gives the dominant contribution to axion production, and the effect of the longitudinal mode photons can be safely neglected. Therefore, we neglect the thermal corrections in the following and calculate the axion production rates by using Eqs.~\eqref{eqn: the production spectra (Primakoff)} and~\eqref{eqn: the production spectra (coalescence)}.

\section{Fluence of photons observed at the Earth}
\label{sec: fluence of photon observed at the Earth}

In this section, we assess the detectability of photon signals arising from the decay of keV-MeV axions produced inside massive stars. In concrete terms, we (i) construct the stellar models by using the stellar evolution code MESA~\cite{Paxton:2010ji,Paxton:2013pj,Paxton:2015jva,Paxton:2017eie,Paxton:2019lxx}, (ii) compute the production rate of axions inside stars, (iii) estimate the fraction of axions that escape from the stellar plasma without undergoing spontaneous decay, (iv) calculate the flux of $X$-ray/$\gamma$-ray, produced by the decay of axions, at the detector on or near Earth, and (v) examine its observability by current/future telescopes. Sec.~\ref{subsec:stellar models} begins with a brief summary of the stellar evolution, followed by a discussion of our choice of target stars as axion laboratories and the methodology for modeling those stars using the MESA stellar evolution code. Estimations of the fraction of axions undergoing spontaneous decay outside stars and subsequent $X$-ray/$\gamma$ signal at the detector are provided in Sec.~\ref{subsec:photon signal at the Earth}. Sec.~\ref{subsec:telescopes} provides information on telescopes and discusses the potential astrophysical backgrounds.

\subsection{Stellar models}
\label{subsec:stellar models} 

\begin{figure*}
    \centering
    \begin{subfigure}{.3\textwidth}
      \centering
      \includegraphics[width=1\linewidth]{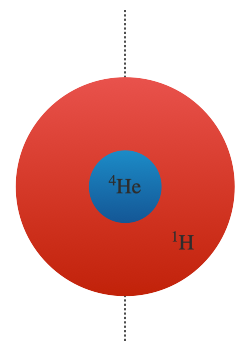}
    \end{subfigure}%
    \begin{subfigure}{.7\textwidth}
      \centering
      \includegraphics[width=1\linewidth]{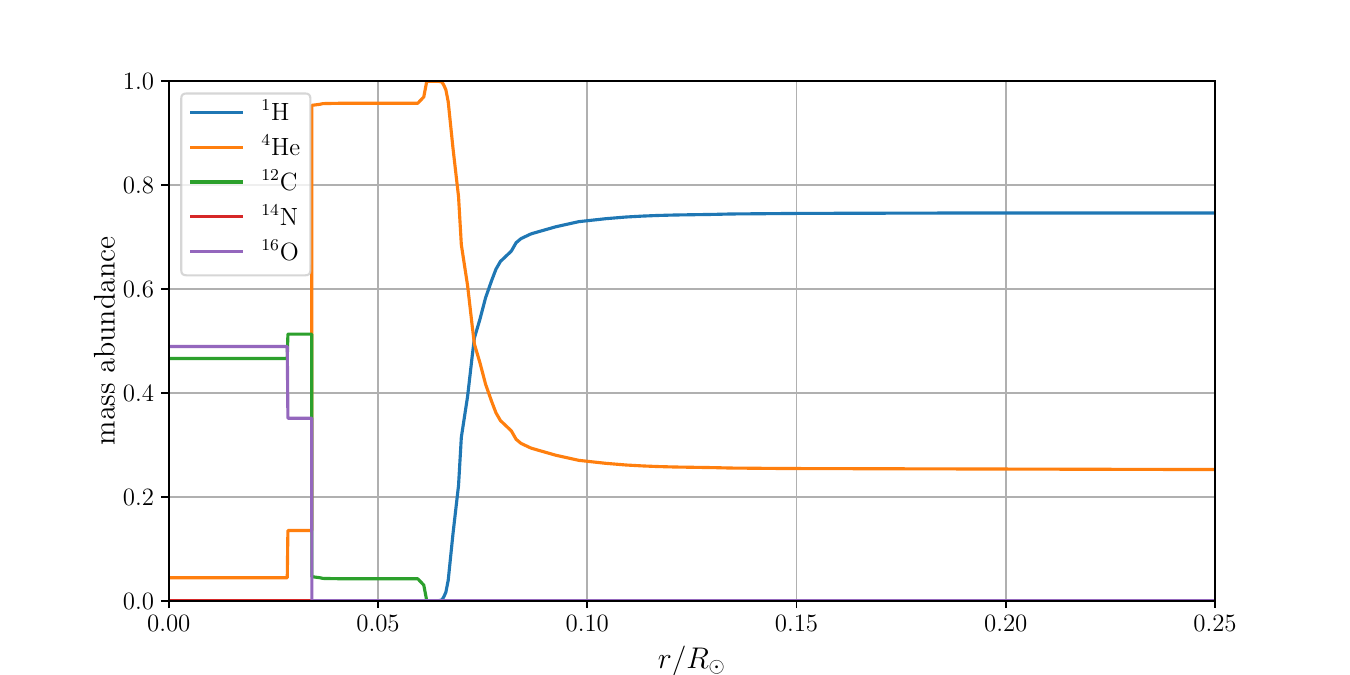}
    \end{subfigure}
    \caption{\label{fig:HB_diagram} Left: The schematic diagram of a HB star. It consists of a helium core and hydrogen shell at the onset of the horizontal branch phase. Heavier nuclei such as $^{12}$C and $^{16}$O are produced via helium burning subsequently. Right: The mass abundance of each nucleus as a function of the distance from the center of the star at the moment when the helium abundance at the center drops to $5\%$. The initial mass, initial helium abundance, and initial metallicity are chosen as $(M_\mathrm{init}, Y, Z) = (0.83M_\odot, 0.254, 0.0005)$. 
    } 
\end{figure*}

\begin{figure*}
    \centering
    \begin{subfigure}{.3\textwidth}
      \centering
      \includegraphics[width=1\linewidth]{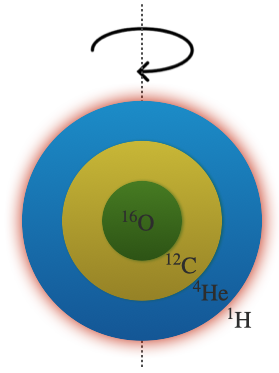}
    \end{subfigure}%
    \begin{subfigure}{.7\textwidth}
      \centering
      \includegraphics[width=1\linewidth]{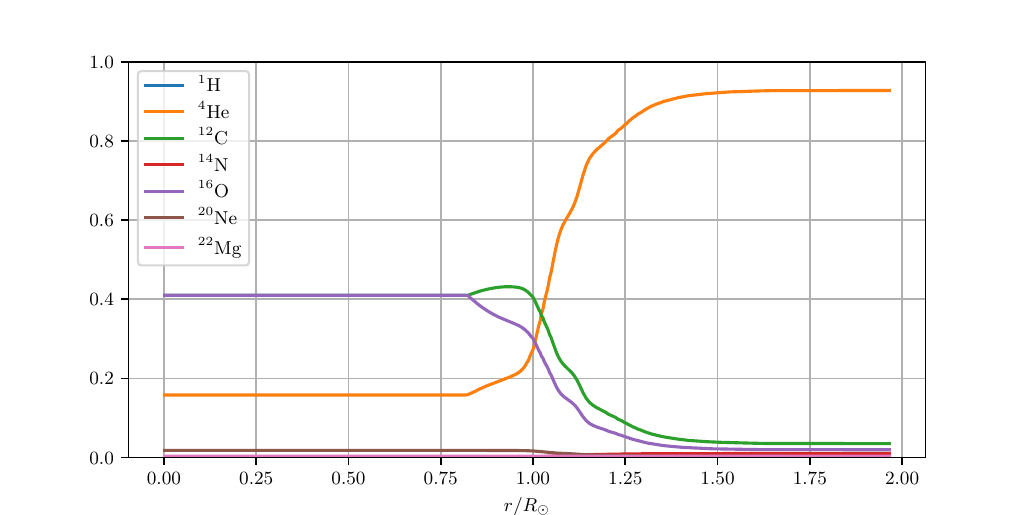}
    \end{subfigure}
    \caption{\label{fig:stars_diagram} Left: The schematic diagram of a WR star. The WR star consists of helium 4 and heavier nuclei (e.g. $^{12}$C, $^{16}$O). The hydrogen envelope is almost stripped away due to its fast rotation. Right: The mass abundance of each nucleus as a function of the distance from the center of the star at the onset of the WC phase. The initial mass, initial rotational velocity, and initial metallicity are chosen as $(M_\mathrm{init}, v_\mathrm{rot}, Z) = (90M_\odot, 180\si{\ km/s}, 0.018)$.} 
\end{figure*}

\subsubsection{History of stellar evolution} 
\label{subsubsec:stellar evolution} 

Formation of protostars takes place in the dense region of the molecular cloud in the interstellar medium. This dense region experiences gravitational contraction, leading to an increase in temperature and pressure. Hydrogen is ignited once the temperature is reached high enough, and consequently, the gravitational contraction force is balanced by the outward pressure from nuclear fusion. These hydrogen-burning stars occupy the main sequence in the Hertzsprung–Russell diagram. MS stars primarily consist of hydrogen at the point of their formation. They generate energy by burning hydrogen in the core either via the pp-chain reaction in less massive MS stars or CNO-cycle in massive MS stars. These processes heat MS stars up to $\mathcal{O}(\mathrm{keV})$ at the core.

MS stars become red-giant branch (RGB) stars when they use up hydrogen fuel in the core. This transition takes place $\mathcal{O}(10^{8-10}~\mathrm{yrs})$ after low- or intermediate- mass stars are born. The core of RGB stars is mainly made of ${}^4\mathrm{He}$, which is a product of the pp-chain and the CNO-cycle. Their cores are inert, and as a result, they contract and are heated. A surrounding shell of hydrogen continues to undergo fusion, leading to the expansion of RGB stars to $\mathcal{O}(100 R_\odot)$. 

In case of low-mass stars ($M_\mathrm{init} \lesssim 2 M_\odot$), degeneracy pressure of a helium core ends up dominating over thermal pressure, i.e. a helium core becomes degenerate. A core keeps contracting by releasing its gravitational potential energy, increasing a core temperature until helium is ignited. Ignition of helium leads to the helium flush, a release of enormous energy in a very short time. High-mass stars ($M_\mathrm{init} \gtrsim 2 M_\odot$) undergo instead a gradual emission of energy since its high core temperature prevents a degenerate helium core, leading to a less intense dynamics. Stars burning helium in their core are called HB stars whose temperature and radius are typically $\mathcal{O}(10)$ keV and $\sim 20 R_\odot$, respectively. Fig.~\ref{fig:HB_diagram} illustrates the composition of a HB star at the end of the HB phase. 

Helium burning yields $^{12}\mathrm{C}$, which subsequently produces $^{16}\mathrm{O}$. Those reactions cease once helium in the core is depleted. Following this, the core made of carbon and oxygen starts contracting while hydrogen and helium shells are expanded to $\mathcal{O}(100) R_\odot$. These stars, asymptotic giant branch (AGB) stars, are then evolved to white dwarfs, neutron stars, or black holes depending on their initial mass.

Initially superheavy stars ($M_\mathrm{init} \gtrsim 30 M_\odot$) could follow a different evolutionary history and become a rarer class of stars called Wolf-Rayet stars~\cite{crowther2007physical}. WR stars are hot and luminous stars whose hydrogen envelopes are stripped away due to either a wind or a binary interaction. They are categorized into the nitrogen-sequence (WN), the carbon-sequence (WC), and the oxygen-sequence (WO) WR stars by their He, N, C, and O emission lines. The categorization by its surface abundance of nuclei is still not fully established. For simplicity, we assume that WR stars are classified by the surface abundances of relevant nuclei. A WR star enters the WN phase when its surface abundance of hydrogen is reduced to $5$\%, the WC phase when its surface abundance of carbon reaches $2$\%, and the WO phase when its surface abundance of oxygen becomes $2$\% as assumed in Ref.~\cite{Dessert:2020lil}.  Fig.~\ref{fig:stars_diagram} illustrates the composition of WR stars. The right panel plots the mass abundance of each nucleus as a function of the distance from the center of a WR star just entering the WC phase. 

\subsubsection{Target stars as axion laboratories}
\label{subsubsec:target stars} 

\begin{figure*}
    \centering
    \begin{subfigure}{.4\textwidth}
      \centering
      \includegraphics[width=\linewidth]{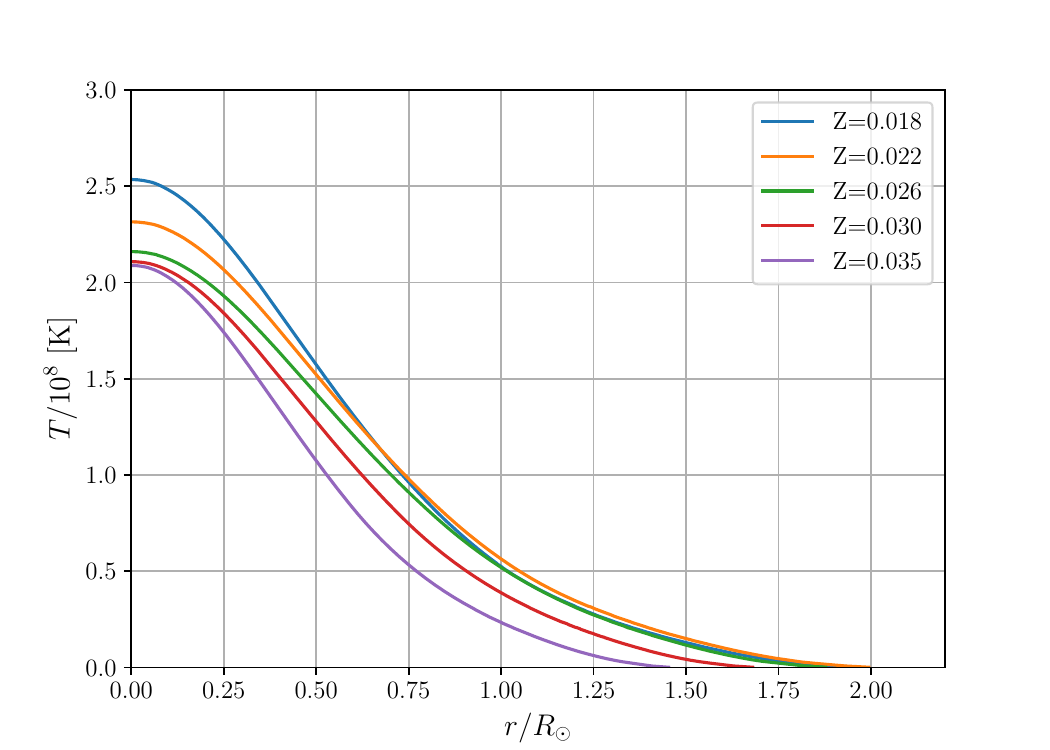}
    \end{subfigure}%
    \begin{subfigure}{.4\textwidth}
      \centering
      \includegraphics[width=\linewidth]{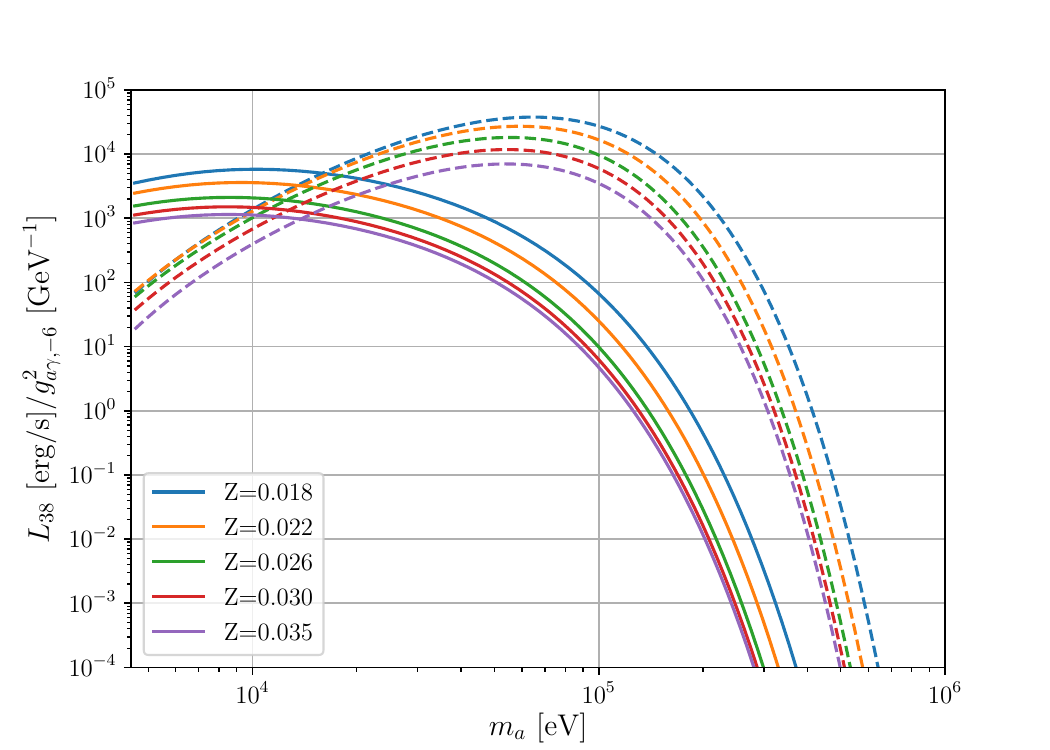}
    \end{subfigure}
    \caption{\label{fig: compare different z} Left: The temperature profile of a WR star at the onset of the WC phase as a function of its radius. Five different metallicity values $Z=0.018, 0.022, 0.026, 0.030$, and $0.035$ are considered.  We set an initial mass and a rotational velocity at the zero-age MS as $85 M_\odot$ and $v = 150 \si{\ km/s}$. Right: Luminosity of axions produced in a WR star with different metallicity by the Primakoff process (solid lines) and photon coalescence (dotted lines). The luminosity is normalized by dividing it by the axion-photon coupling strength $g_{a\gamma} = 10^{-6}$ GeV$^{-1}$.}
\end{figure*}

In this paper, we focus on stars beyond the MS, namely, the HB stars and WR stars, as production sites of axions because they have several appealing features: (i) high temperature of the plasma, $T _{\text{HB}} \sim 10 \si{\ keV}$, is about an order higher than that of MS stars, leading to more efficient production of axions, and (ii) small radius of their photosphere, roughly an order of magnitude smaller than that of RGB stars and AGB stars, resulting in less absorption of photons signal induced by spontaneous decay of axions. 
In the following, we briefly outline the astrophysical objects considered in this study and describe our approach to modeling them by using the MESA stellar evolution code. 

\paragraph{HB stars}

\begin{table}
    \centering
    \small
    \begin{tabular}{|c|cc|} \hline
         & NGC 2808 & NGC 6397 \\ \hline 
         \\[-1em]
         Mass ($M _{\odot}$) & $1.42 \times 10 ^{6}$ & $1.15 \times 10 ^{5}$ \\
         Radius (ly) & $62.8$ & $17.2$ \\
         Distance (kly) & $31.3$ & $7.5$ \\
         Right ascension & $09 ^{\mathrm{h}} 12 ^\mathrm{m} 03$ & $17 ^{\mathrm{h}} 40 ^\mathrm{m} 42$ \\
         Declination & $-64 ^{\circ} 51 ^{'} 48 ^{"}$ & $-53 ^{\circ} 40 ^{'} 27 ^{"}$ \\
         $N_{\rm HB}$ & 1200 & 105 \\ \hline
    \end{tabular}
    \caption{\label{tab: globular clusters}Properties of the two globular clusters used in this study. Data of the radius, distance, and right ascension are taken from Refs.~\cite{harris1996catalog,harris2010new,harris2013catalog}. Masses and the number of HB stars are estimated in Refs.~\cite{boyles2011young} and~\cite{constantino2016treatment,sandquist2000catalogue}, respectively. }
\end{table}

Globular clusters consist of millions of gravitationally bounded stars with roughly the same age and chemical composition. Thus, globular clusters in the right age could host a large number of HB stars. One of the largest Galactic globular clusters is NGC 2808 ($M = 1.42 \times 10 ^{6} M_\odot$), which is located at $d =31.3 \si{\ kly}$ away from us. NGC 6397 ($M = 1.15 \times 10^5 M_\odot$), though about an order of magnitude lighter than NGC 2808, is another promising object because of its vicinity to the Earth with $d=7.5 \si{\ kly}$. Photometry estimates the number of HB stars in each globular cluster as $N _{\rm HB} \sim 1200$ for NGC 2808 and $N _{\rm HB} \sim 105$ for NGC 6397, respectively~\cite{constantino2016treatment,sandquist2000catalogue}. Properties and locations of these globular clusters are summarized in Table~\ref{tab: globular clusters}.

The stellar models are constructed using an open-source 1D stellar evolution code MESA~\cite{Paxton:2010ji,Paxton:2013pj,Paxton:2015jva,Paxton:2017eie,Paxton:2019lxx} version r22.11.1. A model of a HB star is evolved from the pre-MS to the terminal-age core helium burning phase. The initial mass, the helium abundance, and the metallicity of a sample star are set as $M _\mathrm{init} = 0.83 M _{\odot}, Y = 0.254$, and $Z = 0.0005$. For simplicity, we assume that a sample star represents the average property of HB stars in globular clusters. Accordingly, we approximate that the total production rate of axions in each globular cluster as the production rate of a sample star multiplied by $N _{\rm HB}$.

\paragraph{WR stars}

\begin{figure*}
    \centering
    \begin{subfigure}{.4\textwidth}
      \centering
      \includegraphics[width=\linewidth]{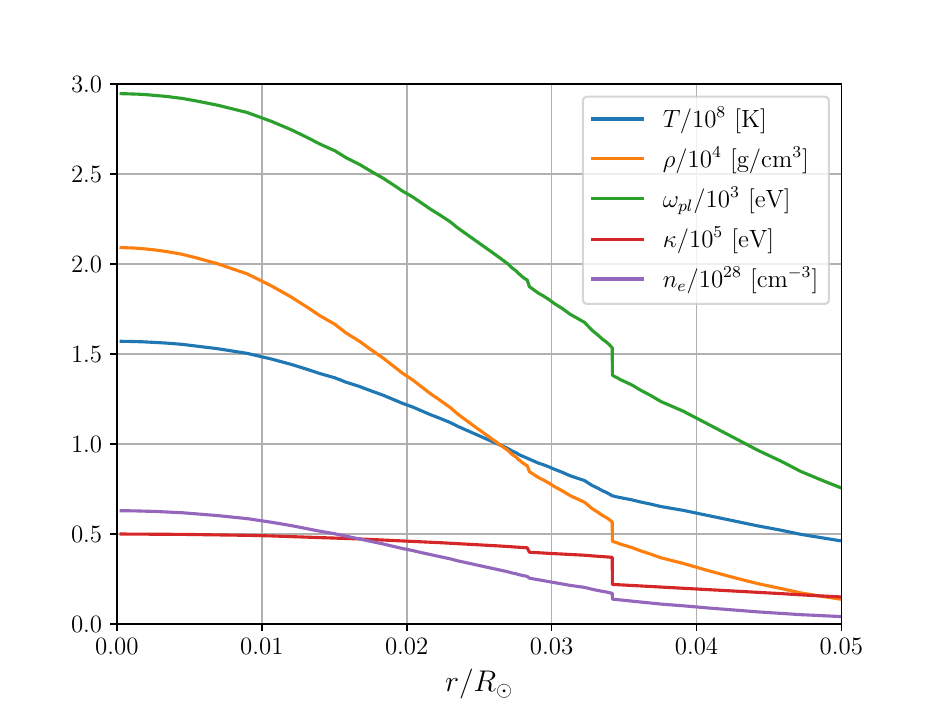}
    \end{subfigure}%
    \begin{subfigure}{.4\textwidth}
      \centering
      \includegraphics[width=\linewidth]{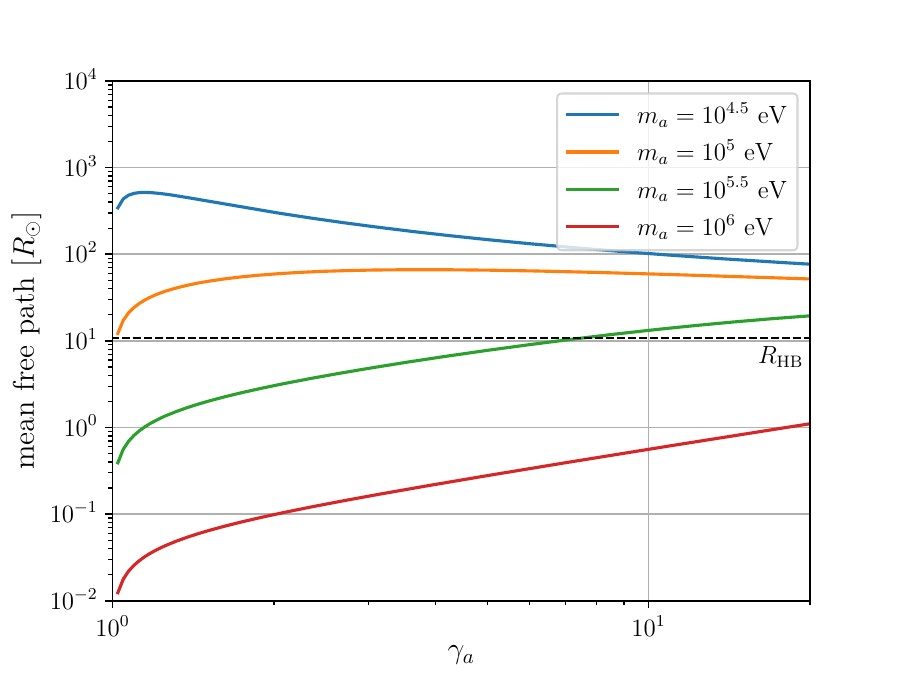}
    \end{subfigure}
  \caption{Left: The profiles of temperature (blue), energy density (orange), plasma frequency (green), the screening scale $\kappa$ (red), and number density of electrons (purple) as a function of the distance from the stellar center. This snapshot is taken when central mass abundance of helium 4 reaches $X _{^{4}\text{He}} = 0.6$. The most internal $0.05 R_\odot$ or $0.44 M_\odot$ regions are plotted. Right: The mean free path of an axion with a mass, $m_a = 10^{4.5}, 10 ^{5}, 10 ^{5.5},$ and $10 ^{6} \si{\ eV}$, with the Lorentz factor $\gamma _{a} \in [1,20]$ is shown. The mean free path is evaluated with the temperature at the HB star core. The dashed black line represents the radius of a sample HB star $R_\mathrm{HB} = 10.9 R_\odot$.}
  \label{fig: HB star properties}
\end{figure*}

For WR stars, we focus on the Quintuplet cluster. The Quintuplet cluster is one of the most massive young clusters in the Milky Way near the Galactic Center. The spectroscopic survey shows that this cluster contains 71 massive stars among which 13 are in the WC and 1 is in the WN phase. Their ages are estimated to fall within the range $t \in (3.0,3.6)\si{\ Myr}$~\cite{clark2018updated}. Requiring the observed nitrogen abundance of WNh stars in the Arches cluster matches with that estimated by MESA and assuming the Arches cluster and the Quintuplet cluster have the same metallicity, the metallicity of the Quintuplet cluster is estimated to be in the range $Z \in (0.018, 0.035)$. For a given metallicity $Z$, the helium abundance $Y$ is found by the following formula:
\begin{align}
    \label{eqn: helium abundance}
    Y = Y_p + \left(\frac{Y_{\text{protosolar}} - Y_p}{Z _{\text{protosolar}}}\right)Z,
\end{align}
where $Y_p = 0.248$ is the primordial helium abundance estimated from a combination of measurements of the CMB power spectra, lensing, and baryon acoustic oscillation~\cite{Planck:2018vyg}, $Y_{\text{protosolar}} = 0.2703$ is the protosolar helium abundance, and $Z_{\text{protosolar}} = 0.0134$ is the protosolar metallicity~\cite{asplund2009chemical}. Each star in the Quintuplet cluster is evolved from the pre-MS until near a core-collapse by using MESA. 

Fig.~\ref{fig: compare different z} plots the temperature profile (left) and luminosity of axions as a function of the axion mass (right) for a sample WR star with different metallicity values. The higher initial metallicity leads to a more efficient mass loss of stars, resulting in larger luminosity of axions~\cite{Dessert:2020lil}. On the other hand, stars with lower initial metallicity tend to predict a smaller radius at the beginning of the WC phase that allows more axions to escape from their photospheres. Since the photon signal from axion decay has nontrivial dependence on the initial metallicity $Z$, we made stellar models and estimated the detectability of photons both for high ($Z=0.035$) and low ($Z=0.018$) metallicities.

Since the Quintuplet cluster is located near the Galactic center, the distance to the cluster is approximated by the Galactocentric distance of the sun $d _{\rm GC} \sim 8 \si{\ kpc}$~\cite{reid1993distance}. Simulations by MESA revealed that WR stars in the WC phase have a higher temperature and a smaller radius, implying that they dominantly contribute to the photon signal from axion decay. Therefore, we consider the production of axions only in 13 WC stars. 

The sample of 13 WC stars is calculated in the following way. We first randomly chose an initial mass $M _\mathrm{init}$ and a rotational speed $v _\mathrm{rot}$ of a sample star. Those are assumed to follow the Kroupa initial mass function~\cite{kroupa2001variation} and the Gaussian distribution function with a mean velocity $\mu _\mathrm{rot} = 100 \si{\ km/s}$ and a velocity dispersion $\sigma _\mathrm{rot} = 140 \si{\ km/s}$~\cite{Hunter:2007be,brott2011rotating}. The initial mass is assumed to fall within the range $M _\mathrm{init} \in (20,200) M_\odot$. This stellar model is evolved from its pre-MS phase until it enters the WC phase. Repeating this process, we construct $\mathcal{O}(100)$ samples of WC stars. Among these samples, only those that predict the age of a star consistent with the spectral analysis, i.e. in the range of $(3.0,3.6)$ Myr at the onset of the WC phase, are retained in our analysis and the rest are discarded. A detailed description of the choice of parameters in our code can be found in Appendix \ref{appendix: WR stars}.

\begin{figure*}
    \centering
    \begin{subfigure}{.4\textwidth}
      \centering     \includegraphics[width=\linewidth]{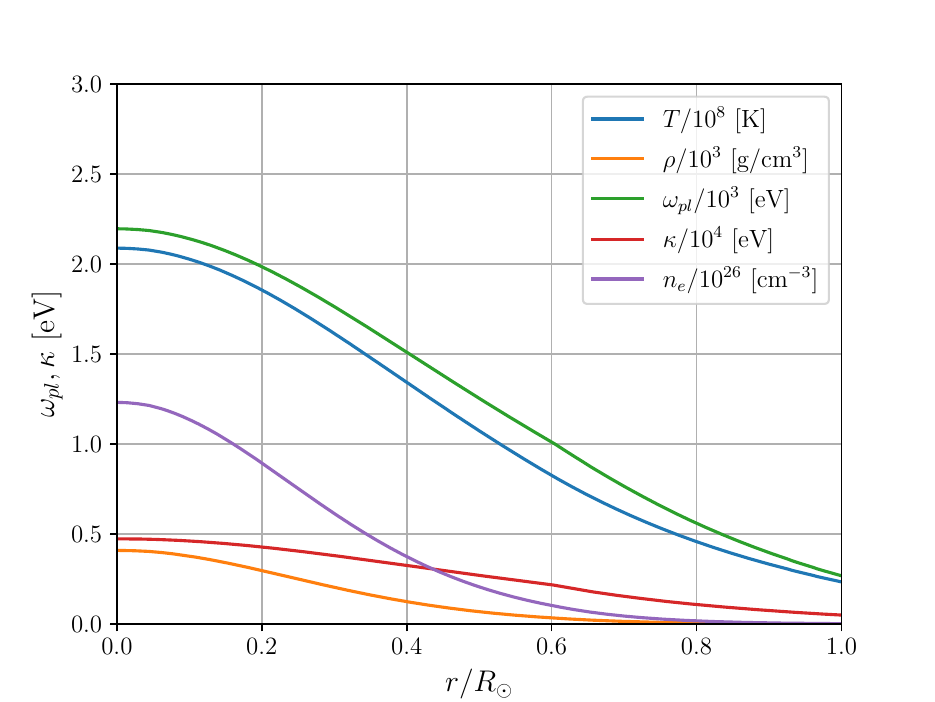}
    \end{subfigure}%
    \begin{subfigure}{.4\textwidth}
      \centering
\includegraphics[width=\linewidth]{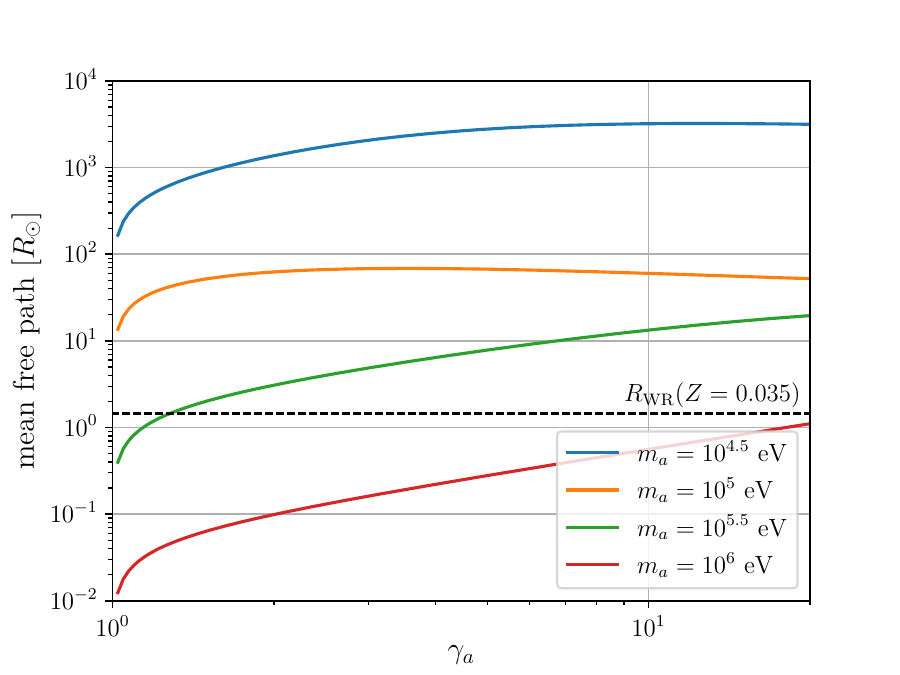}
    \end{subfigure}
  \caption{Left: The profiles of temperature (blue), energy density (orange), plasma frequency (green), the screening scale $\kappa$ (red), and number density of electrons (purple) of a WR star with $Z=0.035$ and an initial mass $85 M_\odot$ when it reaches the WC phase. The most internal $R_\odot$ region is plotted. Right: The mean free path of an axion with several masses with the Lorentz factor $\gamma _{a} \in [1,20]$ at the WR star core. The dashed black line represents the radius of a sample WR star, $R_\mathrm{WR} = 1.45 R_\odot$, for a metallicity $Z=0.035$.}
  \label{fig: WR star properties}
\end{figure*}

The left panel of Figs. \ref{fig: HB star properties} and \ref{fig: WR star properties} show the plasma temperature, density, plasma frequency, and Debye screening length of a sample HB star and WR star as a function of the distance from the stellar center, respectively. The mean free path of axions with several benchmark masses is plotted in the right panels. 

\subsection{Photon signal at the Earth}
\label{subsec:photon signal at the Earth}

\begin{figure*}
    \centering
    \includegraphics[width=150mm]{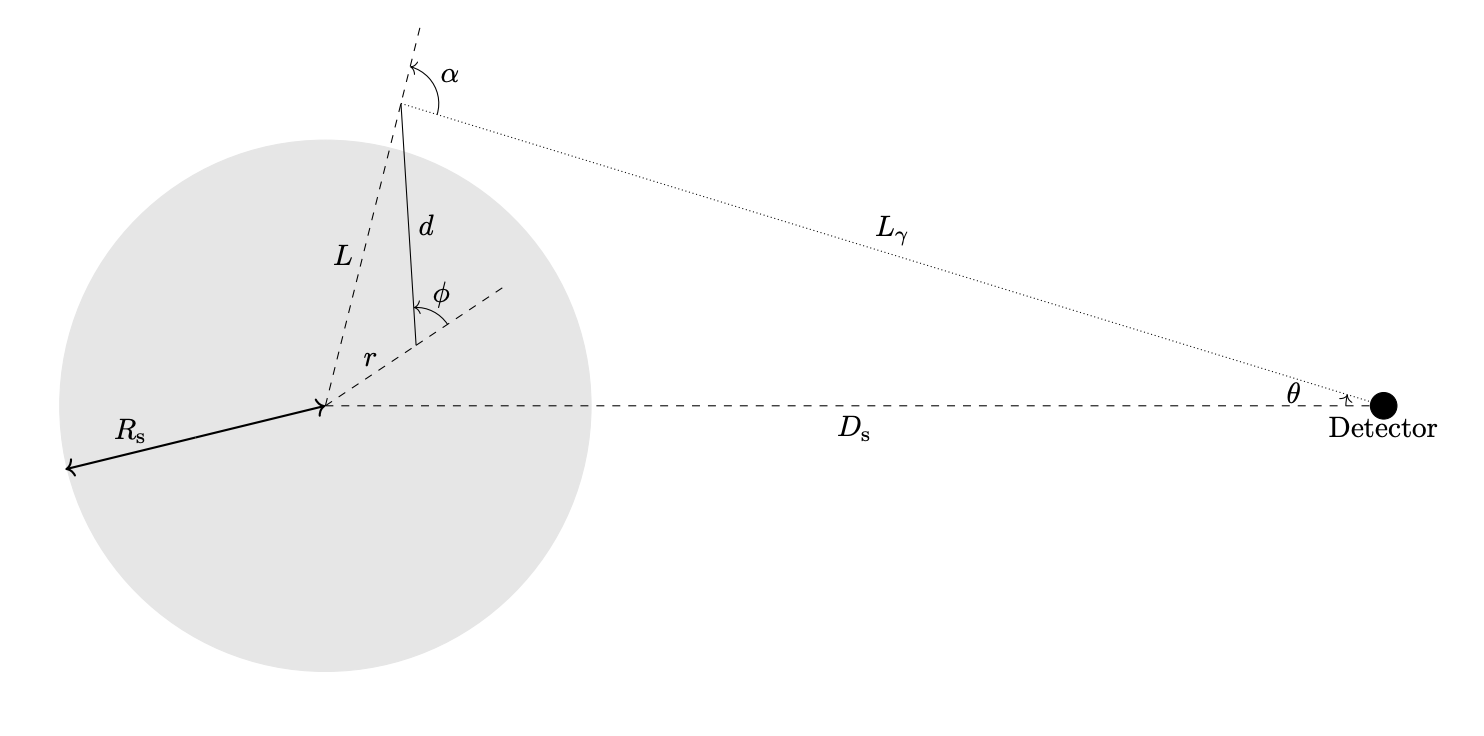}
    \caption{An illustration of the photon detection from spontaneous decay of axions produced inside a star. Axions are produced in the stellar plasma either by the Primakoff process or photon coalescence. Axions that decay outside the photosphere could yield an observable photon signal at the Earth. The radius of the photosphere is denoted by $R _{\mathrm{s}}$ and the distance to the star is $D_s$. 
    }
    \label{fig: geometry of photon detections}
\end{figure*}

Integrating the axion production spectra given in Eqs.~(\ref{eqn: production rate (Primakoff)}) and  (\ref{eqn: production rate (coalescence)}) over the volume  of the star, we arrive at the total axion production rate, i.e. the number of axions produced with energy $E$ per unit time:
\begin{align}
    \label{eqn: axion production rate}
    \frac{\dd^2 N_a}{\dd E \dd t}(t,E)=\int_0^{R_{s}} \dd r 4 \pi r^2 \frac{\dd^2 n_a}{\dd t \dd E}(r, t, E).
\end{align}
where $R_s$ is the stellar radius. 

An axion could undergo the reconversion into a photon in presence of a magnetic field or spontaneously decay into two photons inside a star. These photons are subsequently absorbed in the plasma and therefore do not contribute to an observable signal. In order to incorporate this effect, we rewrite Eq. (\ref{eqn: axion production rate}) as
\begin{align}
    \label{eqn: spectral fluence}
    \frac{\dd^2 N_a}{\dd E \dd t}(E) = \int_0^{R_{s}} \dd r 4 \pi r^2 \frac{\dd^2 n_a}{\dd t \dd E}(r, E) e^{-\tau(d, E)},
\end{align}
where $\tau$ is the optical depth of axions:
\begin{align}
    \tau(r, \phi, E)=\int_0 ^{d} \dd l\ \lambda^{-1}(l, E),
\end{align}
$\lambda$ is the overall mean free path of axions, and $d$ is the distance that an axion travels before escaping from the star's photosphere.

The overall mean free path can be expressed in terms of the mean free path of each process: $\lambda^{-1} = \lambda^{-1} _{\mathrm{Primakoff}} + \lambda^{-1} _{\mathrm{Coalescence}}$.  Each mean free path is obtained by 
\begin{align}
\lambda _{\mathrm{i}} = \beta _{a} \Gamma _{\mathrm{i}} ^{-1} = \frac{\beta _{a} ^{2} E ^{2}}{2 \pi ^{2} e ^{E/T}} \left(\frac{\dd ^{2} N _{a}}{\dd E \dd t}\bigg|_i\right) ^{-1}.
\end{align}
Here $\beta_a = \sqrt{1-(m_a/E) ^{2}}$ is the axion speed. 
The right panels of Figures~\ref{fig: HB star properties} and \ref{fig: WR star properties} show the overall mean free path  for several choices of the axion mass. These plots imply that heavy axions have shorter mean free paths. For comparison, we also show by horizontal dashed lines the typical size of the star; only the axions with a mean free path larger than this value can escape the star. 

Once the axions escape the photosphere of the star and decay into photons, a fraction of them might be detected on the Earth. Fig. \ref{fig: geometry of photon detections} illustrates the geometry of detection of photons from the decay of axions produced in the star (see e.g. Refs.~\cite{Balaji:2022noj, Dev:2023hax} for a detailed discussion of the geometry effect). The distance $L$ the axion travels before decaying can be expressed as a function of the radius of the photosphere $R_s$, the radial distance $r$ of the point where an axion is produced, and the inclination angle $\phi$ of the axion momentum vector relative to the radial direction:
\begin{align}
    L = - r \cos\phi + \sqrt{R _{\mathrm{s}} ^{2} - r ^{2} \sin ^{2} \phi}.
\end{align}

The differential of the fluence $F _{\gamma}$ of photons, i.e., the number of photons per unit time per unit area, from axion decay is given by~\cite{Jaeckel:2017tud,Ferreira:2022xlw,Jaffe:1995sw}
\begin{alignlast}
    \mathrm{d} F_{\gamma} &= 2 \cdot \mathrm{BR}_{a \rightarrow \gamma \gamma} \cdot \frac{1}{4 \pi D_s^{2}} \cdot \frac{\dd^2 N_a }{\dd E \dd t} \dd E \cdot f_{c_{\alpha}}\left(E, c_{\alpha}\right) \dd c_{\alpha} \\
    &\quad \times \frac{\exp \left[-L / l_{a}(E)\right]}{l_{a}(E)} \dd L \cdot \Theta_{\text{cons.}}\left(E, c_{\alpha}, L\right),
    \label{eq:fluence}
\end{alignlast}
where the different factors are explained as follows: (i) The factor 2 takes into account the number of photons produced in each decay of an axion. (ii) $\mathrm{BR}_{a \rightarrow \gamma \gamma}$ is the branching ratio of the decay of an axion into two photons. Since we neglect all couplings except the axion-photon coupling, this branching ratio is set as $1$. (iii) $\dfrac{1}{4 \pi D_s^{2}}$ is a geometric factor for the decrease in luminosity, assuming an isotropic production of axions. (iv) $\dfrac{\dd^2 N_a}{\dd E \dd t} \dd E$ is the number of axions that escape from the stellar photosphere per unit time. (v) $f_{c_{\alpha}}\left(E, c_{\alpha}\right)$ is the distribution of the angle $\alpha$:
    \begin{align}
        f_{c_\alpha}\left(E, c_\alpha\right)=\frac{m_a^2}{2 E^2\left(1-c_\alpha \beta_a\right)^2},
    \end{align}
    as a function of the axion mass, energy and velocity. Here $c_\alpha \equiv \cos \alpha$. (vi) $\dfrac{\exp \left[-L / l_{a}(E)\right]}{l_{a}(E)} \mathrm{d} L$ is the probability for an axion to decay within a distance $[L, L + \mathrm{d}L]$ from the source. The decay length of an axion $l_{a}$ is given by 
    \begin{align}
        l_a(E) = \frac{\beta_a \gamma_a}{\Gamma_0^{a \rightarrow \gamma \gamma}},
    \end{align}
    where $\Gamma_0^{a \rightarrow \gamma \gamma} = g_{a\gamma}^2 m_a^3/64\pi$ is the spontaneous decay rate of an axion in the rest frame. (vii)  $\Theta_{\text{cons.}}\left(E, c_{\alpha}, L\right)$ is the constraint placed on the energy of an axion $E$, the angle $\alpha$, and the distance $L$ so that one of the photons from the decay of an axion arrives at the detector on the Earth. This constraint is given by a product of the Heaviside functions:
    \begin{align*}
        \label{eqn: detection constraints}
        \Theta_{\text{cons.}}\left(E, c_\alpha, L\right)=&\ \Theta\left(E_{\max }-\omega_\gamma\left(\omega, c_\alpha\right)\right) \\
        &\times \Theta\left(\omega_\gamma\left(\omega, c_\alpha\right)-E_{\min }\right) \\
        &\times \Theta\left(c_\theta\left(c_\alpha, L\right)\right) \\
        &\times \Theta\left(c_\theta\left(c_\alpha, L\right)-c_\alpha\right)\\
        &\times \Theta\left(L-R_{s}\right), \atag
    \end{align*}
    where $\omega_\gamma$ is the energy of the photon observed at the Earth, $c _{\theta} = \cos\theta$, and $L _{\gamma}$ is the distance that a photon has to travel until it arrives at the detector. These are given by
    \begin{align}
        \omega_\gamma\left(E, c_\alpha\right)&=\frac{m_a^2}{2 E\left(1-c_\alpha \beta_a\right)}, \\
        c_\theta\left(c_\alpha, L\right)&=\frac{L_\gamma^2+D_s^2-L^2}{2 L_\gamma D_s},\\
        L_\gamma &=L\left(\sqrt{\frac{D_s^2}{L^2}-1+c_\alpha^2}-c_\alpha\right).
    \end{align}
    Each constraint in Eq. (\ref{eqn: detection constraints}) is described below:
    \begin{itemize}
        \item $\Theta\left(E_{\max }-\omega_\gamma\left(\omega, c_\alpha\right)\right) \Theta\left(\omega_\gamma\left(\omega, c_\alpha\right)-E_{\min }\right)$:  The energy of a photon is in the range of each energy band $[E _{\mathrm{min}}, E_{{\mathrm{max}}}]$ of the detector.
        \item $\Theta\left(c_\theta\left(c_\alpha, L\right)\right)$: The photon reaches the detector without being shielded by the Earth.
        \item $\Theta\left(c_\theta\left(c_\alpha, L\right)-c_\alpha\right)$: The parameters geometrically allow a photon to arrive at the detector.
        \item $\Theta\left(L-R_{s}\right)$: The axion decays beyond the photosphere with a radius $R _{s}$.
    \end{itemize}
    The differential fluence in Eq.~\eqref{eq:fluence} is integrated over energy and angle $\alpha$ subject to the constraints given above to obtain the total fluence of the photon signal induced by axion decay, which is then compared with the experimental sensitivity for a given telescope to derive the ALP sensitivity curves in the next section.    

\subsection{Telescopes}
\label{subsec:telescopes}
\begin{table*}[htb!]
{\footnotesize
\centering
\begin{tabular}{cccccc} \hline
           &             & Angular &  &   &  \\
           & Sensitivity & Resolution & Mission & Effective & Observation  \\
   Mission & Range & (at Energy) & Status & Area & Time \\ \hline
    XMM-Newton~\cite{XMM:2001haf} & 0.1-15 keV & 12$^{''}$ (2-10 keV) & 1999-present & $\sim$3000 cm$^2$ & $10^4\si{\ s}$ \\
    NuSTAR~\cite{NuSTAR:2013yza} & 5 keV - 80 keV & 18$^{''}$ & 2012-present & $\sim$1000 cm$^2$  & $10^6 \si{\ s}$\\
    INTEGRAL IBIS/ISGRI~\cite{Lebrun:2003aa} & 15 keV - 1 MeV & 12$^{'}$ & 2002-2023 & 250 cm$^2$ & $10^5 \si{\ s}$ \\
    INTEGRAL IBIS/PICsIT~\cite{DiCocco:2003gr} & 170 keV - 10 MeV & 12$^{'}$ & 2002-2023 & $\sim$1400 cm$^2$ &$10^5 \si{\ s}$ \\
    INTEGRAL SPI~\cite{vedrenne2003spi} & 20 keV - 8 MeV & 2.5$^{\circ}$ & 2002-2023 & $\sim$3000\, cm$^2$ & $10^6 \si{\ s}$ \\
    INTEGRAL JEM-X~\cite{lund2003jem} & 3 keV - 35 keV  & 3$^{'}$ & 2002-2023 & 400 cm$^2$ & $10^6 \si{\ s}$ \\
    SWIFT (BAT)~\cite{SWIFT:2005ngz}  & 15-150 keV &  22$^{'}$ &2004-present & 5200 cm$^2$ (15 keV) & 0.7$\times$19 yrs \\
    eROSITA~\cite{eROSITA:2012lfj} & 0.2-10 keV & 35$^{''}$ (2-8 keV)  & 2019-present & 1500\,cm${}^2$ & $10^5$ s\\
     Insight-HXMT/HE~\cite{Insight-HXMTTeam:2019dqg} & 20 keV - 250 keV & 6$^{'}$ & 2017-present & 4096 cm$^2$ & $\sim$10$^5$ s \\  
    COSI~\cite{Tomsick:2023aue} & 200 keV - 5 MeV &  $\sim$4$^\circ$ (1 MeV) & 2027 (planned) & $\sim 300$\,cm$^2$ & 2 yrs \\
    AMEGO (Compton)~\cite{AMEGO:2019gny} & 200 keV - 10 MeV & $\sim 4^\circ$ (1 MeV)  & Concept & $\sim$300 cm$^2$ & 0.24$\times$ 5 yrs \\
     AMEGO-X~\cite{Caputo:2022xpx} & 100 keV - 1 GeV & $\sim$10$^\circ$ & Concept & 3500 cm$^2$ & 3 yrs \\
    APT (Compton)~\cite{APT:2021lhj} & 200 keV - 10 MeV & $\sim 5^\circ$ (1 MeV) & Concept & 10,000 cm$^2$ & $0.82\times$ 5 yrs \\
  ASTROGAM~\cite{DeAngelis:2021esn} & 100 keV - 1 GeV & $\sim1.5^\circ$ & Concept & 1000 cm$^2$ & 1 yr \\
    e-ASTROGAM~\cite{e-ASTROGAM:2017pxr} & 300 keV - 3 GeV &  $\sim$2$^\circ$ & Concept & 9025 cm$^2$ & 1 yr \\ 
    GECCO~\cite{Orlando:2021get} & 100 keV - 10 MeV & 2$^\circ$ & Concept & 2000 cm$^2$ & $10^6 \si{\ s}$  \\ \hline
\end{tabular}
\caption{\label{tab: specs of telescopes}Summary of properties of the telescopes considered here.   Note that for some gamma-ray instruments, we give the performance in the relevant Compton (not pair-production) regime.}}
\end{table*}
The energy of axions produced in HB and WR stars is comparable either to the temperature of their plasma (for $T \gtrsim m_a$) or the rest mass of axions (for $T \lesssim m_a$). Those axions then spontaneously decay into two photons. As shown in Fig.~\ref{fig: compare different z} right panel, the spectrum of photons is expected to have a peak at $E_\gamma \sim \mathcal{O}(10-100) \si{\ keV}$, and thus, a hard $X$-ray telescope or MeV gamma-ray telescope is best suited for observing those photons.  

We estimated the observability of photons by several existing hard $X$-ray telescope missions XMM-Newton~\cite{XMM:2001haf}, NuSTAR~\cite{NuSTAR:2013yza} and INTEGRAL (which includes IBIS/ISGRI~\cite{Lebrun:2003aa}, IBIS/PICsIT~\cite{DiCocco:2003gr}, SPI~\cite{vedrenne2003spi}, and JEM-X~\cite{lund2003jem}), SWIFT~\cite{SWIFT:2005ngz}, eROSITA~\cite{eROSITA:2012lfj}, Insight-HXMT/HE~\cite{Insight-HXMTTeam:2019dqg}, and one future $X$-ray mission COSI~\cite{Tomsick:2023aue}, as well as several future soft gamma-ray telescope missions AMEGO/AMEGO-X~\cite{AMEGO:2019gny, Caputo:2022xpx},  ASTROGAM/e-ASTROGAM~\cite{Rando:2019fzq, DeAngelis:2021esn}), GECCO~\cite{Orlando:2021get} and APT~\cite{APT:2021lhj}. 
The sensitivity range of each telescope is summarized in Table~\ref{tab: specs of telescopes}. In general, the angular resolution of telescopes depends on the energy of incoming photons. Moreover, a telescope could be operated in more than one mode each of which has a different angular resolution; for example, the angular resolution of GECCO is $\sim 1^{'}$ in the Mask mode while it is $4^\circ - 8^\circ$ in the Compton mode. For simplicity, we assume the angular resolution is independent of the energy of injected photons and is the value listed in  Table~\ref{tab: specs of telescopes}. In order to estimate the detectability of photons, we used the $3\sigma$ continuum sensitivity of the telescopes~\cite{Lucchetta:2022nrm} by assuming the observation times summarized in the Table. We found that INTEGRAL SPI gives the best sensitivity in the relevant ALP parameter space of our interest. This is mainly because of its energy sensitivity range which covers the peak ALP emission spectrum from HB stars and WR stars, as well as its large angular resolution which ensures more photon collection per pixel from the ALP decay region.   

\section{Detectability of photon signals}
\label{sec:results}

\begin{figure}
    \centering
    \includegraphics[width=\linewidth]{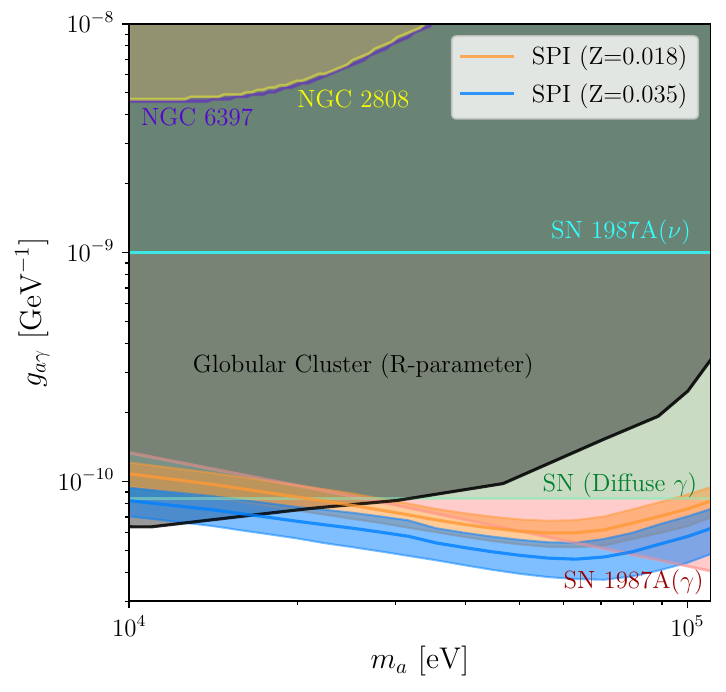}
    \caption{\label{fig: constraints} SPI sensitivity of the photon signal induced by decaying ALPs produced from HB stars located in NGC 2808 (yellow) and NGC 6397 (purple), as well as from WR stars located in the Quintuplet cluster for two different metallicity values of $Z=0.018$ (orange) and $Z=0.035$ (blue). The existing constraints from $R$-parameter measurement in globular clusters, SN 1987A and diffuse photon background from supernovae are shown as shaded regions.     
    }
\end{figure}

In this section, we present our results on the detection of photons resulting from the spontaneous decay of axions produced inside HB and WR stars. Fig.~\ref{fig: constraints} highlights the regions in the $m_a-g_{a\gamma}$ parameter space that can be probed by INTEGRAL SPI. We compare our result to the current limits from the number counts of horizontal branch stars and red-giant branch stars in globular clusters~\cite{Ayala:2014pea,Carenza:2020zil,Lucente:2022wai} (black), the non-observation of photons from the spontaneous decay of axions produced inside SN 1987A (red)~\cite{Ferreira:2022xlw,Hoof:2022xbe,Muller:2023vjm}, anomalous cooling of SN 1987A due to axion emission (cyan)~\cite{Caputo:2022mah}, and diffuse gamma-ray background from decays of axions produced inside supernovae (green)~\cite{Caputo:2022mah}. We find that our result for the HB stars in NGC 2808 and NGC 6397 are two orders of magnitude weaker than the current best constraint in the ${\cal O}(10)$ keV mass range. On the other hand, for the WR stars in the Quintuplet cluster, we obtain better sensitivity which can cover currently unconstrained ALP parameter space in the 10-100 keV region, probing the ALP-photon coupling down to $g_{a\gamma}\gtrsim 4\times 10^{-11}~{\rm GeV}^{-1}$. Note that the WR stars are roughly an order of magnitude smaller in size compared to the HB stars, although both have similar core temperatures. This allows axions with smaller mean free path to escape the photosphere of WR stars, thus increasing the signal sensitivity, as compared to the HB star case. 

We estimated the WR signal detectability for two different initial stellar metallicity: $Z=0.018$ (blue) and $Z=0.035$ (cyan). For each stellar model simulated by MESA, we calculate the photon fluence at the Earth, assuming that all 13 WC stars in the Quintuplet cluster follow the same density, temperature, and abundance profiles. By comparing this fluence to the sensitivity of telescopes, we identify the region in the $m_a-g_{a\gamma}$ parameter space that could be probed by the current/future telescopes. These analyses are repeated for all stellar models. Fig.~\ref{fig: constraints} illustrates the mean of $g_{a\gamma}$ bounds and its $1\sigma$ uncertainty band for two different metallicity values of  $Z=0.018$ and $Z=0.035$ with the SPI detector. The uncertainty band accounts for the variation in stellar profile parameters, such as temperature, density, plasma frequency, etc. The bound itself is derived by requiring that the expected fluence of ALP-induced photon signal in any given pixel of the SPI detector should not exceed the quoted $3\sigma$ sensitivity of the detector. We repeated the same exercise for the other telescopes listed in Table~\ref{tab: specs of telescopes}, but the result for SPI turned out to be the best (for reasons mentioned in Sec.~\ref{subsec:telescopes}); therefore, we do not show the sensitivity curves for other telescopes in Fig.~\ref{fig: constraints}. 

In the regime of large $m_a$ and $g_{a\gamma}$, axion production becomes more efficient, with a greater proportion of axions decaying within the photosphere. On the other hand, in the domain of small $m_a$ and $g_{a\gamma}$,  a smaller number of axions is produced while axions have a longer decay length, (i) facilitating them to escape from the star, (ii) making the expected photon signal more diffuse, and (iii) leading to a smaller fraction of axions that have decayed until they reach the Earth. These factors collectively result in distinct shapes in the constrained regions of the parameter space.

The detectability of the axion-induced photon signal also depends on the potential astrophysical backgrounds. A diffuse background of keV-MeV photons could be emitted by several mechanisms. One of the constituents of $\mathcal{O}$(MeV) gamma-ray background is the decay of a neutral pion produced in astrophysical environments including solar flares, interstellar medium, supernova remnants, molecular clouds, and galaxy clusters~\cite{1987ApJS...63..721M,2004A&A...413...17P,2012MNRAS.421L.102O}. Their subsequent decay forms the diffuse gamma-ray background with energies $E_\gamma \simeq m_{\pi^0}/2$. The diffuse $\mathcal{O}$(keV) photon background is predominantly shaped by thermal emission from white dwarfs, bremsstrahlung, and inverse Compton scattering. Cosmic-ray $e^\pm$ emits photons either by upscattering background low-energy photons (inverse Compton scattering) or traveling in the external electromagnetic fields in which their trajectories are deflected (bremsstrahlung). In this study, we assume the background brightness as described in Ref.~\cite{vedrenne2003spi} to determine the $3\sigma$ sensitivity shown in Fig.~\ref{fig: constraints}.  

As a sanity check, we also made quick estimations on the background $\mathcal{O}$(keV) and $\mathcal{O}$(MeV) photon flux as follows: We used {\tt ximage} to obtain the counts of photons at R7 band ($E \in [1.05,2.04]$ keV) observed by ROSAT~\cite{1995ApJ...454..643S,1997ApJ...485..125S}, and then converted it to the flux by using the {\tt WebPIMMS}~\cite{1994ApJ...424..714S}, confirming that the flux is not high enough to mask the photon signals from decays of axions in this energy band. 

Since the size of each pixel in the ROSAT All-Sky Survey diffuse background maps ($12^{'}$) is larger than the angular extension of the Quintuplet cluster ($\sim 1^{'}$), we just choose the one pixel to obtain the photon count rate $1 \times 10^{-4}\si{\ s^{-1}}$ from the direction of the cluster. This count rate turns out to be equivalent to the flux $1 \times 10^{-14} \si{\ erg/cm^2/s}$ in the same energy band. In order to obtain the flux, we assume that the HI column density is $2.9 \times 10^{20} \si{\ cm^{-2}}$ and the photon spectrum follows the thermal bremsstrahlung spectrum with $kT = 0.3 \si{\ keV}$.

\section{Discussion and Conclusion}
\label{sec:conclusion}

In this article, we investigated the heavy axion production inside massive stars, such as the WR stars in the Quintuplet cluster and the HB stars in two globular clusters, and the detectability of photons from the ALP decay by current and future $X$-ray and gamma-ray telescopes. We find that among the existing and planned telescopes in the keV-MeV sensitivity range, the INTEGRAL SPI is best suited for the purpose of the ALP decay-induced photon signal and can probe the axion-photon coupling down to $g_{a\gamma} \gtrsim 4 \times 10^{-11}\si{\ GeV^{-1}}$ for axion mass $m_a \sim \mathcal{O}(10-100)\si{\ keV}$ in the WR star case, surpassing the existing astrophysical constraints by a whisker. On the other hand, our result for the HB star case turns out to be weaker than the existing constraints.  

Apart from the uncertainties in the WR star profile and elemental composition, a major source of uncertainty in our result is the initial metallicity of WR stars. Therefore, we showed our results for two different values of the initial metallicity within its uncertainty range ($0.018 \leq Z \leq 0.035$), but found that both metallicity values predict comparable luminosity of ALPs (Fig.~\ref{fig: compare different z}), and hence, comparable constraints on the axion-photon coupling (Fig.~\ref{fig: constraints}). 

We have considered only the ALP-photon coupling here for both production and decay of the ALPs. In the presence of other couplings of ALPs to the SM particles (e.g. electrons/nucleons), the production rate would be enhanced, and at the same time, the mean free path would be smaller. However, the ALP decay rate would not change, since for the energy range available inside WR stars (up to a few keV), the ALP is unlikely to decay to other SM particles except photons (and possibly neutrinos, if there is an ALP-neutrino coupling). Therefore, we expect our bounds derived here to remain largely intact. 

\section*{Acknowledgments}

We thank Alex Chen, Saurav Das, Thomas Gehrman, Mike Nowak, Tekeba Olbemo, and Kuver Sinha for useful discussions and comments on the draft. JB, BD, FF, and TO are partly supported by the U.S. Department of Energy under grant No.~DE-SC 0017987. TO is also supported by the University Research Association Visiting Scholars Program.

\onecolumngrid
\appendix

\section{Models of WR stars}
\label{appendix: WR stars}

In this section, we explain our code for simulating WR stars by using the MESA stellar evolution code. A sample star is evolved from the pre-MS until near the core-collapse. Firstly, a system is developed up to the point when a radiative core starts to develop with the test suite ``20M\_pre\_ms\_to\_core\_collapse". Secondly, this star is evolved until a zero age main sequence (ZAMS) by using a test suite ``make\_zams\_ultra\_high\_mass". Lastly, we simulate the evolution of this ZAMS star until near the core-collapse by using a code described below. Parameters in the code are chosen based on Ref.~\cite{choi2016mesa}, and are summarized in Table \ref{tab: MESA parameter choices}.

We choose the nuclear network pp\_cno\_extras\_o18\_ne22.net, which includes 26 species: ${}^{1} \text{H}, {}^{2} \text{H}, {}^{3} \text{He}, {}^{4} \text{He},$ ${}^{7} \text{Li}, {}^{7} \text{Be},$ ${}^{8} \text{B}, {}^{12} \text{C}, {}^{13} \text{C}, {}^{13} \text{N}, {}^{14} \text{N}, {}^{15} \text{N}, {}^{14} \text{O}, {}^{15} \text{O}, {}^{16} \text{O}, {}^{17} \text{O}, {}^{18} \text{O}, {}^{17} \text{F}, {}^{18} \text{F}, {}^{19} \text{F}, {}^{18} \text{Ne}, {}^{19} \text{Ne}, {}^{20} \text{Ne}, {}^{22} \text{Ne}, {}^{22} \text{Mg}, {}^{24} \text{Mg}$ and reactions involving them. A metallicity, an initial mass, and a rotational velocity at a ZAMS of a sample star are chosen as described in the main text. The observed nitrogen abundance of WNh stars in the Arches cluster prefers the metallicity $Z \in (0.018, 0.035)$, which is anticipated to be close to that of the Quintuplet cluster because both are located near the Galactic Center. We perform simulations for two metallicities $Z=0.018, 0.035$. For each of these $Z$ values, the hydrogen and helium abundances are determined by the relation $X = 1-Y-Z$ and Eq.~(\ref{eqn: helium abundance}), respectively. The relative fractions of metals are set by following Ref.~\cite{Grevesse:1998bj}. The initial masses of stars are assumed to follow the Kroupa initial mass function. The probability distribution of initial rotational velocity is described by the Gaussian distribution with a mean velocity $\mu = 100 \si{\ km/s}$ and a velocity dispersion $\sigma = 140 \si{\ km/s}$. 

The Skye equation of state~\cite{jermyn2021skye} is selected since it is the most suited equation of state for temperature and density of our interest. Regarding the opacity table, we choose the Type 2 opacity, which allows to simulate the system with time-dependent abundances of carbon and oxygen. The temperature at the stellar surface is calculated by exploiting the Eddington approximation:
\begin{align}
    \label{eqn: surface temperature}
    T^4(\tau)=\frac{3}{4} T_{\mathrm{eff}}^4\left(\tau+\frac{2}{3}\right), 
\end{align}
and the pressure at the surface is obtained by
\begin{align}
    \label{eqn: surface pressure}
    P=\frac{\tau g}{\kappa}\left[1+P_0 \frac{\kappa}{\tau} \frac{L}{M} \frac{1}{6 \pi}\right],
\end{align}
where $T _{\mathrm{eff}}$ is the effective temperature, $M$ is the stellar mass, $L$ is its luminosity, $\kappa$ is the opacity, and $\tau$ is the optical depth. We have used the Planck units, i.e.~$c=1$ for the speed of light and $G=1$ for the Newton constant. Note that the second term in Eq.~\eqref{eqn: surface pressure} describes the radiation pressure, which is not negligible for massive stars. We used $P_0 = 2$ in our code. 

In massive stars, interactions of photons with ions lead to outward momentum transfer from photons to gas. Consequently, massive stars lose their mass~\cite{Vink:2001cg}. In order to include this effect, we choose the Dutch scheme as a scheme of winds and set an overall scaling factor $\dot{M} \propto \eta$ as equal to 1. Moreover, a rotation could enhance the mass loss rate of hydrogen burning massive MS stars~\cite{langer1998coupled}; the mass loss rate is described by the relation
\begin{align}
    \label{eqn: rotational mass loss}
    \dot{M}(\Omega)=\dot{M}(0)\left(\frac{1}{1-\Omega / \Omega_{\text {crit }}}\right)^{\xi},
\end{align}
where $\Omega_{\text{crit}}$ is the critical angular velocity at the surface of a star and $\xi = 0.43$.

We include mixing processes as follows. Convective mixing is incorporated by using the mixing length theory~\cite{henyey1965studies} for which we need to choose three parameters: the ratio of a mixing length to a pressure scale height $\alpha_{\text{MLT}}$, a multiplicative factor that determines the mean turbulent speed $\nu$, and the efficiency of convection $y$. A convective region is located by the Ledoux criterion. A mixing occurring at convective boundaries, an overshoot mixing~\cite{Herwig:2000sq}, is dealt with independently. An exponentially decaying overshoot mixing is incorporated at all hydrogen burning, helium burning and metal burning convective boundaries with a factor $f = 0.014$. In addition, we include effects of semiconvection~\cite{langer1983semiconvective} and thermohaline mixing~\cite{kippenhahn1980time}. A semiconvection appears in a region that is thermally unstable but stabilized due to a gradient in composition while thermohaline mixing occurs if a thermally stable region is made unstable due to a gradient in composition. Their diffusion coefficients are proportional to the free parameters $\alpha_{\text{sc}}$ and $\alpha _{\text{th}}$. We set $\alpha_{\text{sc}} = 1$ and $\alpha _{\text{th}} = 666$. The rotation of a star induces transports of angular momentum and nuclei. Those transports are approximated by a diffusion process in MESA. We choose parameters that determine the diffusion coefficients as $D _{\text{SH}} = D _{\text{SSI}} = D _{\text{ES}} = D _{\text{GSF}} = D _{\text{ST}} = 1$, $D _{\text{DSI}} = 0$, $f _{c} = 1/30$, and $f _{\mu} = 0.05$ by following Ref.~\cite{heger2000presupernova}.

\begin{table*}
    \centering
    \begin{tabular}{|c|c|} \hline
       Parameter & chosen values  \\ \hline
        nuclear network & pp\_cno\_extras\_o18\_ne22.net  \\
        initial abundances & Eq. (\ref{eqn: helium abundance}), $X = 1-Y-Z$ \\
        fractions of metals & Chosen by following Ref.~\cite{Grevesse:1998bj} \\
        initial mass & Kroupa initial mass function\\
        initial surface rotation & Gaussian distribution, $\mu = 100 \si{\ km/s}$, $\sigma = 140 \si{\ km/s}$ \\
        equation of state & Skye equation of state \\
        opacity & Type 2 \\
        $T$ and $P$ at the surface & Eqs. (\ref{eqn: surface temperature}), (\ref{eqn: surface pressure}), $P _{0}=2$ \\
        wind & Dutch scheme, $\eta = 1$\\
        rotational mass loss & Eq. (\ref{eqn: rotational mass loss}), $\xi = 0.43$ \\
        convection & $\alpha _{\text{MLT}} = 2, \nu = 1/3, y = 8$ \\
        overshoot &  $f_{\text{ov}} = 0.014$\\
        semiconvection & $\alpha_{\text{sc}} = 1$ \\
        thermohaline & $\alpha_{\text{th}} = 666$\\
        rotational mixing & $D _{\text{SH}} = D _{\text{SSI}} = D _{\text{ES}} = D _{\text{GSF}} = D _{\text{ST}} = 1$, $D _{\text{DSI}} = 0$, $f _{c} = 1/30$, $f _{\mu} = 0.05$\\
        \hline 
    \end{tabular}
    \caption{
    \label{tab: MESA parameter choices} A summary of the choice of parameters used in our MESA code.}
\end{table*}

\bibliographystyle{utcaps_mod} 
\bibliography{bibliography}
\end{document}